\documentclass[aps,prd,preprint,nofootinbib,11pt]{revtex4}
\usepackage{amsfonts}
\usepackage{mathrsfs}
\usepackage{graphicx}
\usepackage{amsmath}
\usepackage{amssymb}
\usepackage{subfigure}
\usepackage{epsfig}
\usepackage{graphicx}
\usepackage{ulem}
\usepackage{color}
\usepackage{soul}
\newcommand{\bqa}{\begin{eqnarray}}
\newcommand{\eqa}{\end{eqnarray}}
\newcommand{\beq}{\begin{equation}}
\newcommand{\eeq}{\end{equation}}
\allowdisplaybreaks[1]
\graphicspath{{fig/}{dia/}} \DeclareGraphicsExtensions{.eps}

\begin{document}
\title{\Large Fully strange tetraquark states via QCD sum rules\\[7mm]}

\author{Bing-Dong Wan$^{1,2}$\footnote{wanbd@lnnu.edu.cn} and Ji-Chong Yang$^{1,2}$\footnote{yangjichong@lnnu.edu.cn}\vspace{+3pt}}

\affiliation{$^1$Department of Physics, Liaoning Normal University, Dalian 116029, China\\
$^2$ Center for Theoretical and Experimental High Energy Physics, Liaoning Normal University, Dalian 116029, China}

\author{~\\~\\}

\begin{abstract}
\vspace{0.3cm}
In this paper, we have systematically explored the mass spectrum of fully strange tetraquark candidates within the framework of QCD sum rules, focusing on states with quantum numbers $J^{PC}=0^{++}$, $0^{-+}$, $0^{--}$, $1^{--}$, $1^{+-}$, and $1^{++}$. The analysis reveals the existence of fully strange tetraquark states with masses ranging from approximately $2.07$ to $3.12$ GeV. These predictions are confronted with existing experimental observations of potential fully strange tetraquark resonances, notably the $X(2300)$ recently reported by the BESIII Collaboration, which may be interpreted as a fully strange tetraquark state. 
Furthermore, the possible decay modes of these fully strange tetraquark states are analyzed, providing guidance for their identification in current and future high energy experiments such as BESIII, Belle II, and LHCb. 
\end{abstract}
 \maketitle
\newpage

\section{Introduction}

The study of novel hadronic states--such as multiquarks, hybrids, and glueballs—has gained significant attention in recent years as experimental and theoretical advances continue to challenge the conventional quark model~\cite{GellMann:1964nj,Zweig}. Since the discovery of the $X(3872)$ state~\cite{Choi:2003ue}, more than thirty similar novel  states or candidates have been reported in various experiments. This growing list of observations strongly suggests that many more new hadronic states are likely to be discovered in the near future, marking what can be regarded as a renaissance in hadron spectroscopy. Unraveling the internal structure and underlying dynamics of these newly observed states constitutes one of the most compelling and significant challenges in contemporary hadron physics.

In the light hadron sector, the identification of novel hadronic states remains a significant challenge, primarily due to the small mass splittings between states and the resulting strong mixing among them. This often obscures the distinction between novel and conventional configurations in experimental analyses, except the novel hadronic states possessing exotic quantum numbers, which are forbidden in the conventional quark model, offer cleaner signatures for identifying nontraditional structures. However, with the rapid accumulation of high-precision experimental data in the charm quark sector, the BESIII experiment is now well-positioned to conduct a systematic investigation of hadronic phenomena in this energy region—including, crucially, the search for and study of light novel hadrons~\cite{BESIII:2010gmv,BES:2003aic,BES:2005ega,BESIII:2010vwa,BESIII:2019wkp,BESIII:2016qzq,BESIII:2020vtu,BESIII:2017kqw,BESIII:2017hyw,BESIII:2019cuv}.

Among the various novel hadron candidates, fully strange tetraquark states—composed entirely of two strange quarks and two strange antiquarks ($ss\bar{s}\bar{s}$)—constitute a particularly intriguing and theoretically clean subclass. Owing to their flavor purity, these states are free from mixing with light ($u$, $d$) or heavy ($c$, $b$) quark components, thereby offering a uniquely controlled environment for investigating fundamental aspects of QCD, including quark confinement, color dynamics, and gluon-mediated interactions. Their distinct quark content also enhances their stability against decay into lighter mesons via quark flavor rearrangement. Notably, fully strange tetraquarks can accommodate exotic quantum numbers—such as $J^{PC}=0^{--}$—which are strictly forbidden in conventional quark–antiquark meson configurations. The observation of such states would thus serve as a clear indication of multiquark dynamics beyond the conventional hadron classification scheme.

Recently, the BESIII Collaboration reported the observation of a resonant structure, denoted as $X(2300)$, in a partial wave analysis of the process $\psi(3686)\to\phi\eta\eta^\prime$~\cite{BESIII:2025wpp}. The structure appears prominently in the $\phi\eta$ and $\phi\eta^\prime$ invariant mass spectra, with statistical significances of $9.6~\sigma$ and $5.6~\sigma$, respectively, and the measured decay width of the state is approximately 89 MeV. The measured mass of the $X(2300)$ exhibits a notable discrepancy with the theoretical predictions for conventional strangeonium states as reported in Refs.~\cite{Li:2020xzs,Xiao:2019qhl,Ishida:1986vn,Ebert:2009ub,Oudichhya:2023lva,Wang:2019qyy,Chen:2015iqa}. In contrast, the mass of the fully strange tetraquark state with quantum numbers $J^{PC}=1^{+-}$, as calculated in Ref.~\cite{Liu:2020lpw}, shows good agreement with the observed mass of the $X(2300)$, suggesting a possible tetraquark interpretation. For further studies on fully‑strange tetraquarks, the reader is referred to Refs.~\cite{Su:2022eun,Xi:2023byo,Patel:2025bdu,Dong:2022otb,Dong:2023evc,Ma:2024vsi,Xin:2022qnv}.

Motivated by the observation of the $X(2300)$ resonance, fully strange tetraquark states have attracted renewed attention, especially through the lens of QCD sum rule analyses~\cite{Shifman}, which provide a nonperturbative framework for exploring their possible structure and mass spectrum. QCD sum rules (QCDSR) constitute a QCD-based theoretical framework that systematically incorporates nonperturbative effects. This approach has been successfully applied to a wide range of problems in hadron spectroscopy, providing valuable insights into the structure and properties of conventional and novel hadrons~\cite{Albuquerque:2013ija,Wang:2013vex,Govaerts:1984hc,Reinders:1984sr,P.Col,Narison:1989aq,Tang:2021zti,Qiao:2014vva,Tang:2019nwv,Wan:2019ake,Wan:2020oxt,Wan:2021vny,Wan:2022xkx,Zhang:2022obn,Wan:2022uie,Wan:2023epq,Wan:2024dmi,Tang:2024zvf,Li:2024ctd,Zhao:2023imq,Yin:2021cbb,Yang:2020wkh,Wan:2024fam,Wan:2024pet,Wan:2020fsk,Wan:2024ykm,Tang:2024kmh,Tang:2016pcf,Tang:2015twt,Qiao:2013xca,Qiao:2013raa,Qiao:2013dda,Zhang:2024jvv,Zhang:2023nxl,Zhang:2024ick,Zhang:2024asb,Zhang:2024ulk,Matheus:2006xi,Cui:2011fj,Fu:2018ngx,Huang:2016rro}. The initial step in formulating QCD sum rules involves constructing appropriate interpolating currents that correspond to the hadrons under investigation. These currents encode essential information about the hadrons, such as their quantum numbers and internal structural components. Based on these interpolating currents, the two-point correlation function is defined, which admits two distinct representations: the operator product expansion (OPE) side and the phenomenological side. By matching these two representations through a dispersion relation and applying quark-hadron duality, the QCD sum rules are established. This framework then enables the extraction of hadronic properties, such as the mass spectrum, from the underlying QCD dynamics.

In this work, the masses of the fully strange tetraquark states in molecular configurations with $J^{PC}=0^{++}$, $0^{-+}$, $0^{--}$, $1^{--}$, $1^{+-}$, and $1^{++}$ are investigated within the framework of QCD sum rules. Since the $1^{-+}$ states were investigated in our previous work~\cite{Wan:2022xkx}, and the $0^{+-}$ state does not admit a molecular configuration, these two states are excluded from the present analysis. The organization of the paper is as follows. Following the Introduction, a concise overview of the QCD sum rules framework and the essential formulas employed in our calculations are presented in Sec.\ref{Formalism}. The numerical analysis and corresponding results are discussed in Sec.\ref{Numerical}. The possible tetraquark decay modes are given in Sec. \ref{decay}. Finally, a brief summary and concluding remarks are provided in the last section.

\section{Formalism}\label{Formalism}

To evaluate the mass spectrum of fully strange tetraquark states within the QCD sum rule framework, the initial step is to construct suitable interpolating currents. The procedure for constructing these currents is as follows. First, all possible currents are listed, specifying their constituent quark content and corresponding Dirac gamma matrix structures. 

In general, fully strange tetraquark states may possess a rich internal structure. Several types of interpolating currents have been widely discussed in the literature, including molecular (di-meson) currents, diquark–antidiquark currents, and color-adjoint currents. Each construction emphasizes a different possible organization of quarks and probes different components of the underlying QCD dynamics. Consequently, their predictions for the mass spectrum and decay properties may differ quantitatively.

In the present work, we choose to focus on the molecular-type currents of the form $[\bar{s}s][\bar{s}s]$, which are particularly suitable for describing possible meson–meson bound or resonant configurations in the fully strange sector. This choice is further motivated by the proximity of the expected masses to relevant two-meson thresholds and by the widespread use of such currents in previous QCD sum rule studies of exotic states. We emphasize that the results presented in this paper are therefore restricted to the molecular configuration and should be interpreted within this specific framework. A systematic and quantitative comparison among different current constructions, although highly interesting and important, is beyond the scope of the present work and is left for future investigations.

Subsequently, parity ($P$) and charge conjugation ($C$) transformations are applied to these currents to identify their transformation properties. By selecting the currents with the desired quantum numbers  $J^{PC}$ and eliminating redundant currents via Fierz rearrangements, a complete and non-redundant set of interpolating currents is obtained.

Via the aforementioned procedure, the interpolating currents for fully strange tetraquark states with $J^{PC}=0^{++}$ in molecular configurations are constructed in the following forms:
\begin{eqnarray}\label{current0++}
j^A_{0^{++}}(x)&=& [\bar{s}_a(x)\gamma_5 s_a(x)][\bar{s}_b(x) \gamma_5 s_b(x)] \;,\label{Ja0++}\\
j^B_{0^{++}}(x)&=& [\bar{s}_a(x) s_a(x)][\bar{s}_b(x)  s_b(x)] \;,\label{Jb0++}\\
j^C_{0^{++}}(x)&=& [\bar{s}_a(x)\gamma_\mu s_a(x)][\bar{s}_b(x) \gamma_\mu s_b(x)] \;,\label{Jc0++}\\
j^D_{0^{++}}(x)&=&[\bar{s}_a(x) \gamma_\mu \gamma_5 s_a(x)][\bar{s}_b(x) \gamma_\mu \gamma_5 s_b(x)] \;, \label{Jd0++}\\
j^E_{0^{++}}(x)&=& [\bar{s}_a(x)\sigma_{\mu \nu} s_a(x)][\bar{s}_b(x) \sigma_{\mu \nu} s_b(x)] \;,\label{Je0++}
\end{eqnarray}
where the subscripts $a$ and $b$ are color indices.

The interpolating currents for fully strange tetraquark states with $J^{PC}=0^{-+}$ in molecular configurations are constructed in the following forms:
\begin{eqnarray}\label{current0-+}
j^A_{0^{-+}}(x)&=&i [\bar{s}_a(x)\gamma_5 s_a(x)][\bar{s}_b(x) s_b(x)] \;,\label{Ja0-+}\\
j^B_{0^{-+}}(x)&=& i[\bar{s}_a(x) \sigma_{\mu \nu}  s_a(x)][\bar{s}_b(x) \sigma_{\mu \nu} \gamma_5 s_b(x)]\;.\label{Jb0-+}
\end{eqnarray}

The interpolating currents for fully strange tetraquark states with $J^{PC}=0^{--}$ in molecular configurations are constructed in the following forms:
\begin{eqnarray}\label{current0--}
j^A_{0^{--}}(x)&=&i [\bar{s}_a(x) \gamma_\mu \gamma_5 s_a(x)][\bar{s}_b(x) \gamma_\mu s_b(x)] \;.\label{Ja0--}
\end{eqnarray}

The interpolating currents for fully strange tetraquark states with $J^{PC}=1^{--}$ in molecular configurations are constructed in the following forms:
\begin{eqnarray}\label{current1--}
j_{1^{--}}^{A,\;\mu}(x)&=&i [\bar{s}_a(x) s_a(x)][\bar{s}_b(x) \gamma_\mu s_b(x)] \;,\label{Ja1--}\\
j_{1^{--}}^{B,\;\mu}(x)&=& i[\bar{s}_a(x) \sigma_{\mu \nu} \gamma_5  s_a(x)][\bar{s}_b(x) \gamma_\nu \gamma_5 s_b(x)]\;.\label{Jb1--}
\end{eqnarray}

The interpolating currents for fully strange tetraquark states with $J^{PC}=1^{+-}$ in molecular configurations are constructed in the following forms:
\begin{eqnarray}\label{current1+-}
j_{1^{+-}}^{A,\;\mu}(x)&=&i [\bar{s}_a(x)\gamma_5 s_a(x)][\bar{s}_b(x) \gamma_\mu s_b(x)] \;,\label{Ja1+-}\\
j_{1^{+-}}^{B,\;\mu}(x)&=& i[\bar{s}_a(x) \sigma_{\mu \nu}  s_a(x)][\bar{s}_b(x) \gamma_\nu \gamma_5 s_b(x)]\;.\label{Jb1+-}
\end{eqnarray}

The interpolating currents for fully strange tetraquark states with $J^{PC}=1^{++}$ in molecular configurations are constructed in the following forms:
\begin{eqnarray}\label{current1++}
j_{1^{++}}^{A,\;\mu}(x)&=&i [\bar{s}_a(x) s_a(x)][\bar{s}_b(x) \gamma_\mu \gamma_5 s_b(x)] \;,\label{Ja1++}\\
j_{1^{++}}^{B,\;\mu}(x)&=& i[\bar{s}_a(x) \sigma_{\mu \nu} \gamma_5  s_a(x)][\bar{s}_b(x) \gamma_\nu s_b(x)]\;.\label{Jb1++}
\end{eqnarray}

With the currents (\ref{Ja0++})-(\ref{Jb1++}), the two-point correlation function can be readily established, i.e.,
\begin{eqnarray}
\Pi_{J^{PC}}^k(q^2) &=& i \int d^4 x e^{i q \cdot x} \langle 0 | T \{ j_{J^{PC}}^k (x),\;  j_{J^{PC}}^k (0)^\dagger \} |0 \rangle \;,\\
\Pi_{J^{PC},\;\mu\nu}^k(q^2) &=& i \int d^4 x e^{i q \cdot x} \langle 0 | T \{ j_{J^{JC}}^{k,\;\mu} (x),\;  j_{J^{JC}}^{k,\;\nu} (0)^\dagger \} |0 \rangle \;.
\end{eqnarray}
Here, $\Pi(q^2)$ and $\Pi_{\mu\nu}(q^2)$ denote the correlation functions corresponding to states with spin $J=0$ and $1$, respectively, the index $k$ runs from $A$ to $E$ for $0^{++}$ states, takes the value $A$ for $0^{--}$ state,  and runs from $A$ to $B$ in other case, and $ |0 \rangle$ denotes the physical vacuum. The correlation function $\Pi_{\mu\nu}(q^2)$ corresponding to spin $J=1$ states can be expressed in the following Lorentz-covariant form:
\begin{eqnarray}
\Pi_{\mu\nu}(q^2) &=&-\Big( g_{\mu \nu} - \frac{q_\mu q_\nu}{q^2}\Big) \Pi_1(q^2)+ \frac{q_\mu q_\nu}{q^2}\Pi_0(q^2)\;,
\end{eqnarray}
where the subscripts $1$ and $0$, respectively, denote the quantum numbers of the spin 1 and 0 mesons.

On the phenomenological side, the correlation function $\Pi(q^2)$ can be represented as a dispersion integral over the physical spectrum, with the ground-state tetraquark contribution explicitly isolated, i.e.,
\begin{eqnarray}
\Pi_{J^{PC}}^{phen,\;k}(q^2) & = & \frac{(\lambda_{J^{PC}}^k)^2}{(M_{J^{PC}}^k)^2 - q^2} + \frac{1}{\pi} \int_{s_0}^\infty d s \frac{\rho_{J^{PC}}^k(s)}{s - q^2} \; , \label{hadron}
\end{eqnarray}
where $M$ denotes the mass of the tetraquark state, $\lambda$ is the coupling constant of the hadron, and $\rho(s)$ represents the spectral density, which encapsulates the contributions from higher excited states and the continuum above the threshold $s_0$.

In the OPE representation, the correlation function $\Pi(q^2)$ can be expressed through a dispersion relation as follows:
 \begin{eqnarray}
\Pi^{OPE,\;k}_{J^{PC}} (q^2) = \int_{s_{min}}^{\infty} d s
\frac{\rho^{OPE,\;k}_{J^{PC}}(s)}{s - q^2}+\Pi_{J^{PC}}^{sum,\;k}(q^2)\; .
\label{OPE-hadron}
\end{eqnarray}
Here, $s_{min}$ denotes the kinematic threshold, which typically corresponds to the squared sum of the current-quark masses of the hadron \cite{Albuquerque:2013ija}. $\Pi^{sum}$ represents the contributions to the correlation function that do not possess an imaginary part but yield nontrivial terms after the Borel transformation. The spectral density is given by $\rho^{OPE}(s) = \text{Im} [\Pi^{OPE}(s)] / \pi$, and
\begin{eqnarray}
\rho^{OPE}(s) & = & \rho^{pert}(s) + \rho^{\langle \bar{s} s
\rangle}(s) +\rho^{\langle G^2 \rangle}(s) + \rho^{\langle \bar{s} G s \rangle}(s)\nonumber\\
&+& \rho^{\langle \bar{s} s \rangle^2}(s)
+ \rho^{\langle G^2 \rangle\langle \bar{s} s \rangle}(s)
+ \rho^{\langle \bar{s} s \rangle\langle \bar{s} G s \rangle}(s) \;. \label{rho-OPE}
\end{eqnarray}

\begin{figure}
\includegraphics[width=8.8cm]{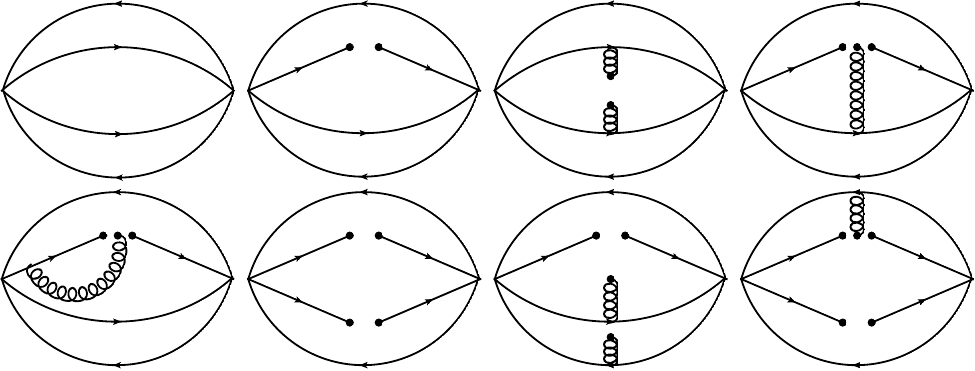}
\caption{The typical Feynman diagrams related to the correlation function, where the solid lines stand for the quarks and the spiral ones for gluons.} \label{feyndiag}
\end{figure}

To calculate the spectral density on the OPE side, as given in Eq.~(\ref{rho-OPE}), the full light-quark propagator $S_{ij}^q(x)$ is utilized, namely,
\begin{eqnarray}
S^q_{i j}(x) \! \! & = & \! \! \frac{i \delta_{i j} x\!\!\!\slash}{2 \pi^2
x^4} - \frac{\delta_{i j} m_q}{4 \pi^2 x^2} - \frac{i t^a_{i j} G^a_{\alpha\beta}}{2^{5}\; \pi^2 x^2}(\sigma^{\alpha \beta} x\!\!\!\slash
+ x\!\!\!\slash \sigma^{\alpha \beta}) - \frac{\delta_{i j}}{12} \langle\bar{q} q \rangle + \frac{i\delta_{i j}
x\!\!\!\slash}{48} m_q \langle \bar{q}q \rangle - \frac{\delta_{i j} x^2}{192} \langle g_s \bar{q} \sigma \cdot G q \rangle \nonumber \\ &+& \frac{i \delta_{i j} x^2 x\!\!\!\slash}{2^7\times 3^2\;} m_q \langle g_s \bar{q} \sigma \cdot G q \rangle - \frac{t^a_{i j} \sigma_{\alpha \beta}}{192}
\langle g_s \bar{q} \sigma \cdot G q \rangle
+ \frac{i t^a_{i j}}{768} (\sigma_{\alpha \beta} x \!\!\!\slash + x\!\!\!\slash \sigma_{\alpha \beta}) m_q \langle
g_s \bar{q} \sigma \cdot G q \rangle \;,
\end{eqnarray}
where the vacuum condensates are explicitly exhibited. For a more detailed discussion of the quark propagator, readers are referred to Refs.~\cite{Wang:2013vex, Albuquerque:2013ija}. The Feynman diagrams corresponding to the various terms in Eq.(\ref{rho-OPE}) are schematically illustrated in Fig.~\ref{feyndiag}.

By applying the Borel transformation to Eqs.(\ref{hadron}) and(\ref{OPE-hadron}), and equating the OPE representation with the phenomenological side of the correlation function $\Pi(q^2)$, one can derive the mass of the tetraquark state as:
\begin{eqnarray}
M_{J^{PC}}^k(s_0, M_B^2) &=& \sqrt{- \frac{L_{J^{PC},\;1}^k(s_0, M_B^2)}{L_{J^{PC},\;0}^k(s_0, M_B^2)}} \;, \\ \label{mass-Eq}
\lambda_{J^{PC}}^k(s_0, M_B^2)&=&\sqrt{Exp(\frac{M_{J^{PC}}^k(s_0, M_B^2)^2}{M_B^2} )\;L_{J^{PC},\;0}^k(s_0, M_B^2)}\;.\label{lambda-Eq}
\end{eqnarray}
Here $L_0$ and $L_1$ are respectively defined as
\begin{eqnarray}
L_{J^{PC},\;0}^k(s_0, M_B^2) =  \int_{s_{min}}^{s_0} d s \; \rho^{OPE,\;k}_{J^{PC}}(s) e^{-
s / M_B^2} +\Pi^{sum,\;k}_{J^{PC}}(M_B^2)  \;,  \label{L0}
\end{eqnarray}
and
\begin{eqnarray}
L_{J^{PC},\;1}^k(s_0, M_B^2) =
\frac{\partial}{\partial{\frac{1}{M_B^2}}}{L_{J^{PC},\;0}^k(s_0, M_B^2)} \; .
\end{eqnarray}

\section{Numerical analysis}\label{Numerical}
In the numerical analysis, we adopt widely accepted input parameters as reported in Refs.~\cite{Matheus:2006xi, Cui:2011fj,P.Col,Tang:2019nwv,Govaerts:1984hc,Reinders:1984sr}, namely: $m_u=2.16^{+0.49}_{-0.26}\; \text{MeV}$, $m_d=4.67^{+0.48}_{-0.17}\; \text{MeV}$, $m_s=(95\pm5)\; \text{MeV}$, $\langle \bar{q} q \rangle = - (0.23 \pm 0.03)^3 \; \text{GeV}^3$, $\langle \bar{s} s \rangle=(0.8\pm0.1)\langle \bar{q} q \rangle$, $\langle \bar{q} g_s \sigma \cdot G q \rangle = m_0^2 \langle\bar{q} q \rangle$, $\langle \bar{s} g_s \sigma \cdot G s \rangle = m_0^2 \langle\bar{s} s \rangle$, $\langle g_s^2 G^2 \rangle = (0.88\pm0.25) \; \text{GeV}^4$, and $m_0^2 = (0.8 \pm 0.2) \; \text{GeV}^2$.

Moreover, two additional parameters, $s_0$ and $M_B^2$ are introduced in the process of establishing the QCD sum rules. These parameters are determined following the standard procedure by satisfying two well-established criteria\cite{Shifman,Reinders:1984sr,P.Col,Albuquerque:2013ija,Govaerts:1984hc}. The first criterion concerns the convergence of the operator product expansion (OPE), which is ensured by examining the relative contributions of higher-dimensional condensates to the total OPE. A suitable Borel window for $M_B^2$ is selected such that the OPE remains convergent in the chosen region. The second criterion requires that the pole contribution (PC) from the ground state should dominate over the continuum, typically accounting for more than $50\%$ of the total spectral density~\cite{Shifman,Reinders:1984sr,Govaerts:1984hc}. These two criteria can be mathematically expressed as:
\begin{eqnarray}
  R^{OPE}= \left| \frac{L_{0}^{dim=8}(s_0, M_B^2)}{L_{0}(s_0, M_B^2)}\right|\, ,
\end{eqnarray}
\begin{eqnarray}
  R^{PC} = \frac{L_{0}(s_0, M_B^2)}{L_{0}(\infty, M_B^2)} \; . \label{RatioPC}
\end{eqnarray}

To determine an appropriate value for the continuum threshold $s_0$, we follow a procedure analogous to that employed in Refs.~\cite{Reinders:1984sr, P.Col, Albuquerque:2013ija}. In this approach, the goal is to identify an optimal value of $s_0$ that yields a stable Borel window for the extracted mass of the fully strange tetraquark state. Within this window, the mass prediction should exhibit minimal dependence on the Borel parameter $M_B^2$, ensuring the reliability of the QCD sum rule analysis. In practical calculations, $\sqrt{s_0}$ is varied by $0.1$ GeV around a central value to determine the corresponding lower and upper bounds, and hence the uncertainties of $\sqrt{s_0}$ \cite{Albuquerque:2013ija}.

\begin{figure}
\includegraphics[width=6.8cm]{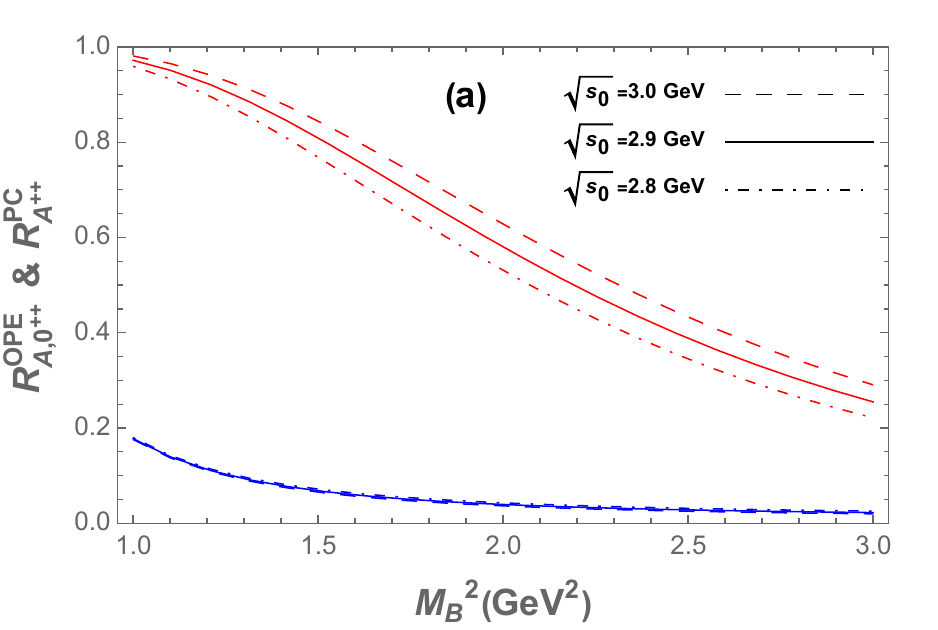}
\includegraphics[width=6.8cm]{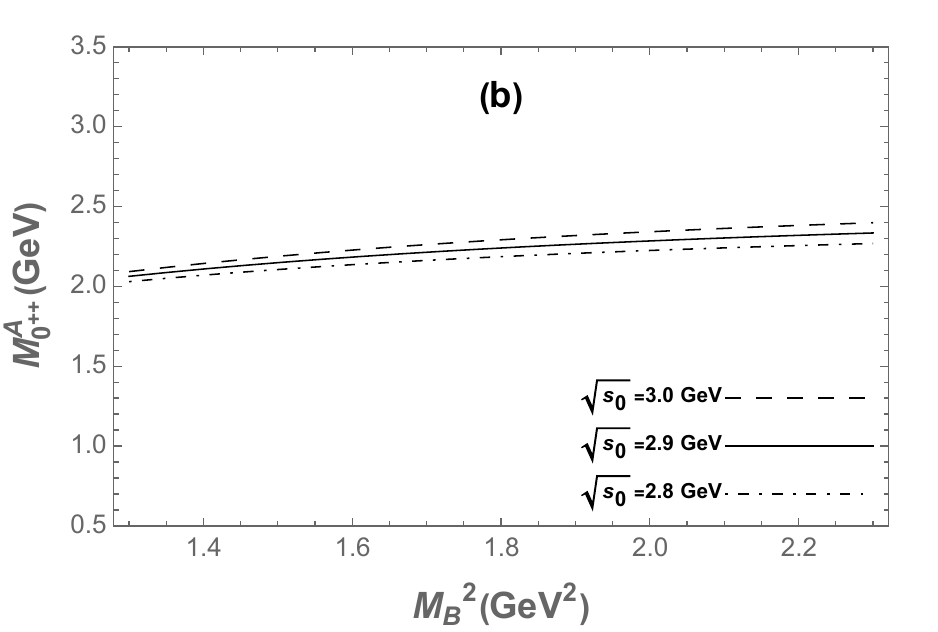}
\caption{ (a) The ratios ${R^{OPE}_{A,0^{++}}}$ and ${R^{PC}_{A,0^{++}}}$ as functions of the Borel parameter $M_B^2$ for different values of $\sqrt{s_0}$ for current (\ref{Ja0++}), where blue lines represent ${R^{OPE}_{A,0^{++}}}$ and red lines denote ${R^{PC}_{A,0^{++}}}$. (b) The mass of $0^{++}$ fully strange tetraquark state as a function of the Borel parameter $M_B^2$ for different values of $\sqrt{s_0}$.} \label{figA0++}
\end{figure}

With the aforementioned preparations, the mass spectrum of the light tetraquark states can now be evaluated numerically. For the current $J_{0^{++}}^A$, as an example, the OPE convergence ratio $R^{OPE}_{A,\;0^{++}}$ and the pole contribution ratio $R^{PC}_{A,\;0^{++}}$ are plotted as functions of the Borel paramete $M_B^2$ in Fig.\ref{figA0++}(a), for three different values of the continuum threshold $\sqrt{s_0}=2.8$, $2.9$, and $3.0$ GeV. The dependence of the extracted mass $M_{0^{++}}^A$ on the Borel parameter is illustrated in Fig.\ref{figA0++}(b). From these analyses, an optimal Borel window is identified as $1.4 \le M_B^2 \le 2.2\; \text{GeV}^2$, within which both the OPE convergence and pole dominance criteria are satisfied. Accordingly, the mass of the fully strange tetraquark state corresponding to the current $J_{0^{++}}^A$ is extracted to be:
\begin{eqnarray}
M^{A}_{0^{++}} &=& (2.23\pm 0.15)\; \text{GeV}\;,\\\label{mA0++}
\lambda^{A}_{0^{++}} &=& (6.4\pm 0.6)\times10^{-3}\; \text{GeV}^5\;,
\end{eqnarray}

Using the same analysis framework, we evaluate the masses associated with the other interpolating currents. The corresponding operator product expansion (OPE) contributions, pole contributions, and the extracted mass values are summarized in Table~\ref{mass}. The uncertainties in the mass predictions reported in Table~\ref{mass} mainly originate from the variations of the input parameters, including the quark masses, vacuum condensates, and the continuum threshold parameter $\sqrt{s_0}$.

\begin{table}
\begin{center}
\renewcommand\tabcolsep{10pt}
\caption{The continuum thresholds, Borel parameters, predicted masses, and predicted decay constant of fully strange tetraquark states.}\label{mass}
\begin{tabular}{cccccc}\hline\hline
$J^{PC}$       &Current   & $\sqrt{s_0}\;(\text{GeV})$     &$M_B^2\;(\text{GeV}^2)$ &$M^X\;(\text{GeV})$   &$\lambda^X\;(10^{-3}\text{GeV}^5)$\\ \hline
$0^{++}$       &$A$         & $2.9\pm0.1$                             &$1.4-2.2$                      &$2.23\pm0.15$            &$6.4\pm0.6$  \\
                      &$B$        & $2.9\pm0.1$                             &$1.4-2.1$                      &$2.24\pm0.16$            &$5.3\pm0.6$\\
                      &$C$        & $3.1\pm0.1$                             &$1.4-2.1$                      &$2.50\pm0.14$            &$11.0\pm1.2$\\
                      &$D$        & $3.1\pm0.1$                             &$1.4-2.1$                      &$2.66\pm0.14$            &$10.8\pm1.4$\\
                      &$E$        & $3.4\pm0.1$                             &$1.4-2.2$                      &$3.00\pm0.08$            & $33.7\pm5.0$\\ \hline
$0^{-+}$        &$A$         & $3.3\pm0.1$                             &$1.8-2.6$                      &$2.57\pm0.16$            &$7.8\pm0.3$\\
                      &$B$        & $3.5\pm0.1$                             &$1.6-2.4$                      &$3.07\pm0.05$            & $48.9\pm4.0$\\\hline
$0^{--}$         &$A$         & $3.1\pm0.1$                             &$1.4-2.1$                      &$2.46\pm0.13$            &$9.4\pm0.9$\\\hline
$1^{--}$        &$A$         & $2.9\pm0.1$                             &$1.4-2.1$                      &$2.46\pm0.15$            &$4.7\pm0.7$\\
                      &$B$        & $3.1\pm0.1$                             &$1.3-2.0$                      &$2.59\pm0.09$            &$8.7\pm1.5$\\\hline
$1^{+-}$        &$A$         & $2.9\pm0.1$                             &$1.4-2.1$                      &$2.29\pm0.14$           & $3.7\pm0.3$\\
                      &$B$        & $3.1\pm0.1$                             &$1.3-2.0$                      &$2.63\pm0.11$            & $8.2\pm1.6$\\\hline
$1^{++}$        &$A$         & $3.3\pm0.1$                             &$1.5-2.2$                      &$2.72\pm0.14$           &$5.3\pm0.5$\\
                      &$B$        & $3.5\pm0.1$                             &$1.7-2.4$                      &$3.03\pm0.08$            &$10.5\pm0.7$\\
\hline
 \hline
\end{tabular}
\end{center}
\end{table}

\section{Discussion on Mass Predictions and Decay Channels}\label{decay}

\subsection{Interpretation of Multiple Mass Predictions}

It should be emphasized that the multiple mass values obtained for the $0^{++}$ channel do not represent theoretical uncertainties of a single state. Rather, they correspond to different interpolating currents with distinct internal structures. Each current may preferentially couple to a particular configuration of the fully strange tetraquark system, leading to different mass predictions and decay patterns.

As an illustrative example, for the $0^{++}$ states, the structure of the interpolating currents provides insight into their dominant decay channels. Specifically, Eq.~(1) corresponds to a coupling of two $0^{-+}$ components and is therefore more likely to decay into $\eta\eta$; Eq.~(2) corresponds to a coupling of two $0^{++}$ components, favoring the $f_0 f_0$ decay mode; Eq.~(3) corresponds to a coupling of two $1^{--}$ components, with $\phi\phi$ as the dominant decay channel; and Eq.~(4) corresponds to a coupling of two $1^{++}$ components, which preferentially decays into $f_1 f_1$.

It is worth noting that the large mass splittings observed among the predicted $0^{++}$ states are a natural consequence of the internal structures probed by the different interpolating currents. For instance, $j^{B}_{0^{++}}$ couples to two $0^{++}$ mesons, leading to a relatively lower mass, while $j^{E}_{0^{++}}$ couples to two $2^{--}$ mesons, resulting in a significantly higher mass. The $2^{--}$ strange mesons have not yet been observed experimentally, likely due to their high mass. Therefore, the substantial mass difference between $j^B_{0^{++}}$ and $j^E_{0^{++}}$ is physically plausible and reflects the distinct internal structures of the tetraquark states. This point has been explicitly discussed in the revised manuscript to aid interpretation of the mass spectrum.

\subsection{Decay Channels of Fully-Strange Tetraquark States}

To finally ascertain these fully strange tetraquark states, the straightforward procedure is to reconstruct them from their decay products, though the detailed characters still ask for more investigation. The decay properties of fully strange tetraquark states strongly depend on their quantum numbers $J^{PC}$, internal structure, and available phase space. Below, we briefly analyze the dominant and allowed decay channels for each quantum number:
\begin{enumerate}
\item  For $J^{PC}=0^{++}$ states. The scalar channel can decay via S-wave into two pseudoscalar or two vector mesons. Possible dominant decay modes include:    $\phi\phi$, $\eta\eta$, $\eta^\prime\eta^\prime$, $\eta\eta^\prime$, $f_0f_0$, $f_1f_1$ and $K\bar{K}$ channels. These are all OZI-allowed and likely lead to relatively broad widths, depending on the mass and phase space. 
 \item  For $J^{PC}=0^{-+}$ states. Being a pseudoscalar state, its decays typically proceed via P-wave into: $\phi\eta$, $\phi\eta^\prime$, and $K\bar{K^\ast}$ channels. These modes are sensitive to angular momentum and available energy. If the mass is close to threshold, the width may be narrow.
 \item  For $J^{PC}=0^{--}$ states. This is an exotic quantum number not accessible by conventional $q\bar{q}$ mesons. Its decays are constrained and potentially suppressed. Allowed multi-body or radiative decays include: $\phi\eta\pi$ and $\eta\eta\gamma$ channels. Due to its exotic nature and limited phase space, this state may be relatively narrow and easier to identify experimentally.
 \item  For $J^{PC}=1^{--}$ states. This vector state can be directly produced in $e^+e^-$ collisions. Its dominant decays are typically: $\phi\eta$, $\phi\eta^\prime$, $K\bar{K}$, and $K^\ast\bar{K}$ channels.
 \item  For $J^{PC}=1^{+-}$ states. This axial-vector state can decay via: $\phi\eta$, $\phi\eta^\prime$, and $K\bar{K}^\ast$ channels. 
 \item  For $J^{PC}=1^{++}$ states. Similar to the $1^{+-}$ case but with different parity, it can decay into: $\phi\eta$ and $K^\ast\bar{K}$ channels.The partial widths and branching ratios are sensitive to internal structure and coupling strengths.
\end{enumerate}

 The decay channels of fully strange tetraquarks offer clear experimental signatures, especially through final states such as $\phi\phi$, $\phi\eta$, and $K\bar{K}^\ast$. Exotic quantum number states like $0^{--}$ stand out due to their unusual decay modes and suppressed widths, making them key targets for discovery. Future searches at BESIII, Belle II, and LHCb can test these predictions by analyzing invariant mass spectra in strange-rich final states.
 
The dominant and allowed decay channels of fully-strange tetraquark states summarized in Table~\ref{tab:decay_channels}.
 
 \begin{table}[ht]
\centering
\caption{Dominant and allowed decay channels of fully-strange tetraquark states.}
\label{tab:decay_channels}
\begin{tabular}{|c|c|c|}
\hline
$J^{PC}$ & Type & Possible Decay Channels \\
\hline
$0^{++}$ & Scalar & $\phi\phi$, $\eta\eta$, $\eta^\prime\eta^\prime$, $\eta\eta^\prime$, $f_0f_0$, $f_1f_1$, $K\bar{K}$ \\
$0^{-+}$ & Pseudoscalar & $\phi\eta$, $\phi\eta^\prime$, $K\bar{K^\ast}$ \\
$0^{--}$ & Exotic & $\phi\eta\pi$, $\eta\eta\gamma$ \\
$1^{--}$ & Vector & $\phi\eta$, $\phi\eta^\prime$, $K\bar{K}$, $K^\ast\bar{K}$ \\
$1^{+-}$ & Axial-vector & $\phi\eta$, $\phi\eta^\prime$, $K\bar{K}^\ast$ \\
$1^{++}$ & Axial-vector & $\phi\eta$, $K^\ast\bar{K}$ \\
\hline
\end{tabular}
\end{table}

\subsection{Estimates of Relative Branching Ratios}

While a full calculation of decay widths is beyond the scope of the present study, we can provide qualitative estimates of the relative branching ratios of the dominant decay channels using simple phenomenological arguments. These estimates are based on the quantum numbers of the initial states, the allowed partial waves, and available phase space.

For the $0^{++}$ states, S-wave decays into two pseudoscalar or two vector mesons are dominant. Among them, decays to channels such as $\phi\phi$ and $\eta\eta$, $f_0f_0$, and $f_1f_1$ are expected to have relatively large branching ratios due to favorable phase space, CKM suppressed, and OZI-allowed couplings, whereas channels with $K\bar{K}$ may be somewhat suppressed.  

For the $0^{-+}$ states, P-wave decays such as $\phi\eta$, $\phi\eta'$, and $K\bar{K}^\ast$ are allowed. The $\phi\eta$ channel is likely to dominate due to larger phase space and fewer angular momentum barriers.  

For the exotic $0^{--}$ states, the limited phase space and exotic quantum numbers imply that multi-body decays like $\phi\eta\pi$ and radiative decays such as $\eta\eta\gamma$ may have comparable contributions, although the total width is expected to be narrow.  

For the $1^{--}$ and $1^{+-}$ states, vector-pseudoscalar and vector-vector channels such as $\phi\eta$, $\phi\eta'$, $K\bar{K}$, and $K^\ast\bar{K}$ are favored. Among these, channels with lower mass thresholds and allowed S-wave decays are expected to have relatively higher branching ratios.  

For the $1^{++}$ states, decays into $\phi\eta$ and $K^\ast\bar{K}$ are expected, with $\phi\eta$ likely being dominant due to phase space considerations.  

These qualitative estimates, together with the dominant decay channels summarized in Table~\ref{tab:decay_channels}, provide a more quantitative guidance for experimental searches and help to identify the most promising channels for detecting fully strange tetraquark states.

\section{Summary}

In this work, we have systematically investigated the mass spectrum of fully strange tetraquark states ($ss\bar{s}\bar{s}$) in molecular configurations with quantum numbers $J^{PC} = 0^{++},\ 0^{-+},\ 0^{--},\ 1^{--},\ 1^{+-},$ and $1^{++}$ within the framework of QCD sum rules (QCDSR). By constructing appropriate interpolating currents and applying standard QCDSR techniques we obtained mass predictions for each quantum number channel. The resulting masses fall within the range of approximately $2.07$ to $3.12$ GeV, depending on the quantum numbers and the interpolating currents used, which are summarized in Table~\ref{mass}. Notably, the mass of the $J^{PC}=1^{+-}$ state shows good agreement with the $X(2300)$ resonance recently reported by the BESIII Collaboration, hinting at a possible exotic tetraquark interpretation. Furthermore, the possible decay modes of these fully strange tetraquark states have been analyzed in detail.

Among these, particular attention is paid to the exotic $J^{PC} = 0^{--}$ channel, which is forbidden in conventional quark-antiquark meson configurations and thus serves as a strong indicator of multiquark or gluonic degrees of freedom. The observation of a state with such quantum numbers would provide compelling evidence for the existence of non-conventional hadrons and offer a direct window into the exotic sector of QCD.

These findings not only enrich the theoretical understanding of multiquark states but also offer concrete and testable predictions for future experimental searches at BESIII, Belle II, LHCb, and other high energy facilities. Owing to its flavor-pure composition and the absence of mixing with light or heavy quarks, the fully strange tetraquark system provides a particularly clean and ideal environment for probing novel hadronic structures and exploring the nonperturbative dynamics of QCD.

It should be noted that the results in Ref. \cite{Jafarzade:2025qvx} suggest that the $X(2300)$ prefers an $s\bar{s}q\bar{q}$ structure, while the $X(2500)$ and $X(2600)$ are more consistent with an $s\bar{s}s\bar{s}$ configuration. To definitively determine the internal structure of the $X(2300)$, a practical approach is to analyze its various decay products, which are discussed in Sec. \ref{decay} of this paper.

In order to place our mass predictions for fully‑strange tetraquark states into a broader theoretical context, we compare our results with those obtained in recent quark‑model and potential-model studies. For instance, Ref.~\cite{Liu:2020lpw} uses a nonrelativistic potential quark model (without invoking a diquark-antidiquark approximation), and predicts an $ss\bar{s}\bar{s}$ tetraquark spectrum with ground and excited $0^{++}$ masses around $2.2$ GeV and up to $3.3$ GeV. This mass range overlaps with the lower end of our predicted spectrum, in particular, some of our $0^{++}$ candidates lie near their ground‑state region, thus providing a nontrivial cross‑validation of our sum‑rule based analysis.

More recently, fully strange tetraquark resonant states as the cousins of X(6900) \cite{Ma:2024vsi} employed the Gaussian expansion method within a constituent quark potential model to solve the four‑body Schrödinger equation and identify resonant states for the $ss\bar{s}\bar{s}$ system. They obtain a series of resonances and compact states in the mass region of $2.7$-$3.3$ GeV, with widths ranging from sub-MeV to around $50$ MeV. Though their predicted masses are generally higher than our central values, the partial overlap, especially in the excited region, suggests that different modeling assumptions can lead to moderate variations, which is also reflected in our results via different interpolating currents.

The deviations between our sum-rule results and quark‑model predictions can be qualitatively understood as arising from fundamentally different assumptions about the internal structure of the tetraquark states (for instance, molecular‑type vs compact four‑quark configurations), different treatments of interquark interactions and relativistic effects, and methodological limitations inherent to each approach (e.g., truncation of the operator product expansion or basis‑set dependence in potential models). As a result, our work should be viewed as complementary to quark‑model studies: while those capture one class of possible compact/resonant configurations, our sum‑rule analysis — especially when employing different interpolating currents — covers a broader set of structural possibilities, including loosely‑bound molecular–type states.

In sum, we believe that the consistency (partial overlap) between our predicted mass ranges and those from potential models lends credence to the plausibility of fully‑strange tetraquark states, and highlights the importance of considering different theoretical frameworks in parallel. This comparison has been added to the revised manuscript to provide a more complete theoretical context for our results.

%%%%%%%%%%%%%%%%%%%%%%%%%%%%%%%%%%%%%%%%%%%%%%%%%%%%%%%%%%%%%%%%%%%%%%
\vspace{.5cm} {\bf Acknowledgments} \vspace{.5cm}

This work was supported in part by the National Natural Science Foundation of China under Grants 12575106 and 12147214, and Specific Fund of Fundamental Scientific Research Operating Expenses for Undergraduate Universities in Liaoning Province under Grants No. LJ212410165019.

%%%%%%%%%%%%%%%%%%%%%%%%%%%%%%%%%%%%%%%%%%%%%%%%%%%%%%%%%%%%%%%%%%%%%%%

\begin{widetext}
\appendix

\section{The ratios $R^{OPE}$, $R^{PC}$, and the masses $m$ are plopted as functions of Borel parameter $M_B^2$}\label{App_B}

We display the figures of the $R^{OPE}$, $R^{PC}$, and the masses $m$ as functions of Borel parameter $M_B^2$ below.

\begin{figure}
\includegraphics[width=6.8cm]{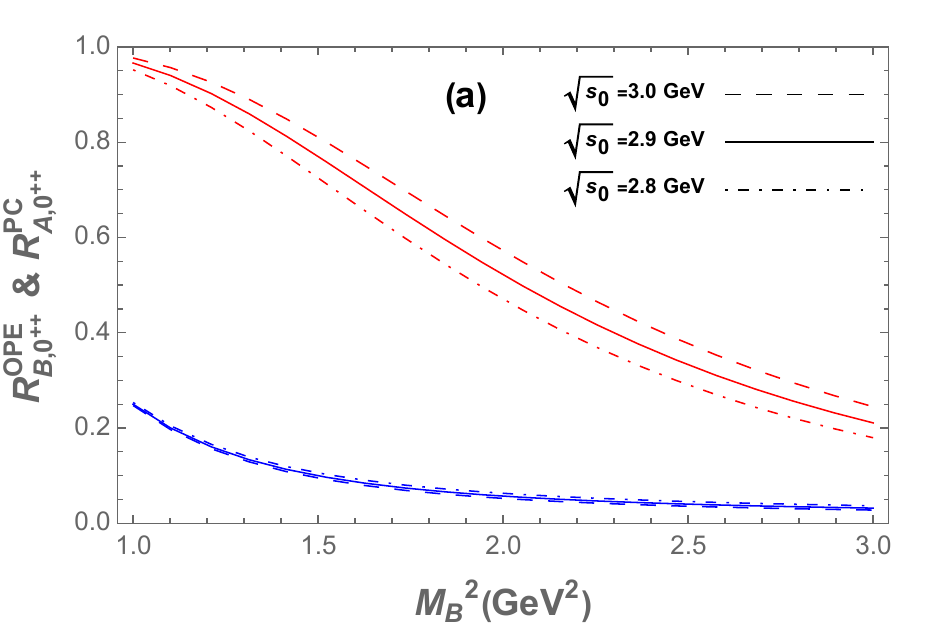}
\includegraphics[width=6.8cm]{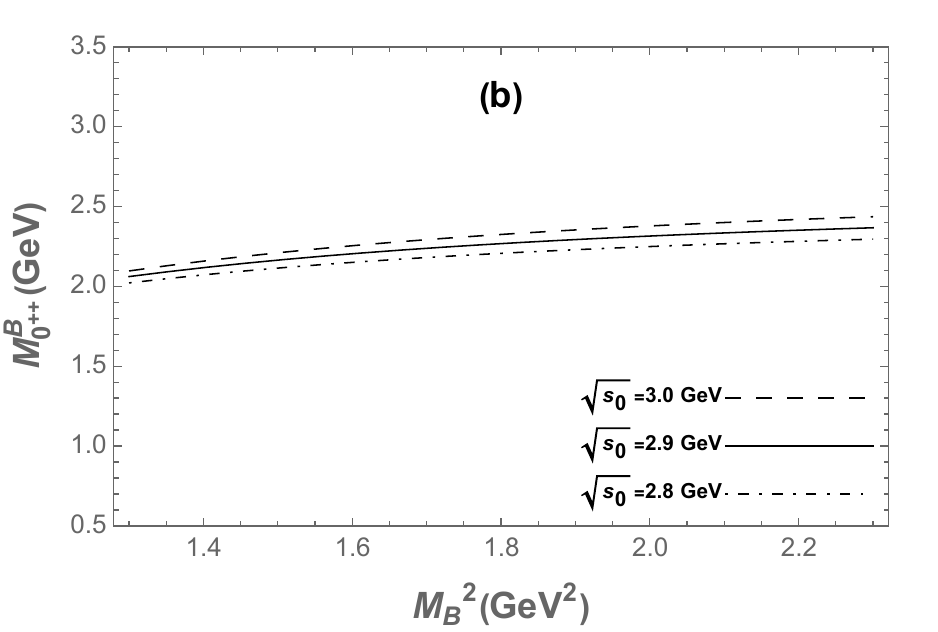}
\caption{ (a) The ratios ${R^{OPE}_{B,0^{++}}}$ and ${R^{PC}_{B,0^{++}}}$ as functions of the Borel parameter $M_B^2$ for different values of $\sqrt{s_0}$ for current (\ref{Jb0++}), where blue lines represent ${R^{OPE}_{B,0^{++}}}$ and red lines denote ${R^{PC}_{B,0^{++}}}$. (b) The mass of $0^{++}$ fully strange tetraquark state as a function of the Borel parameter $M_B^2$ for different values of $\sqrt{s_0}$.} \label{figB0++}
\end{figure}

\begin{figure}
\includegraphics[width=6.8cm]{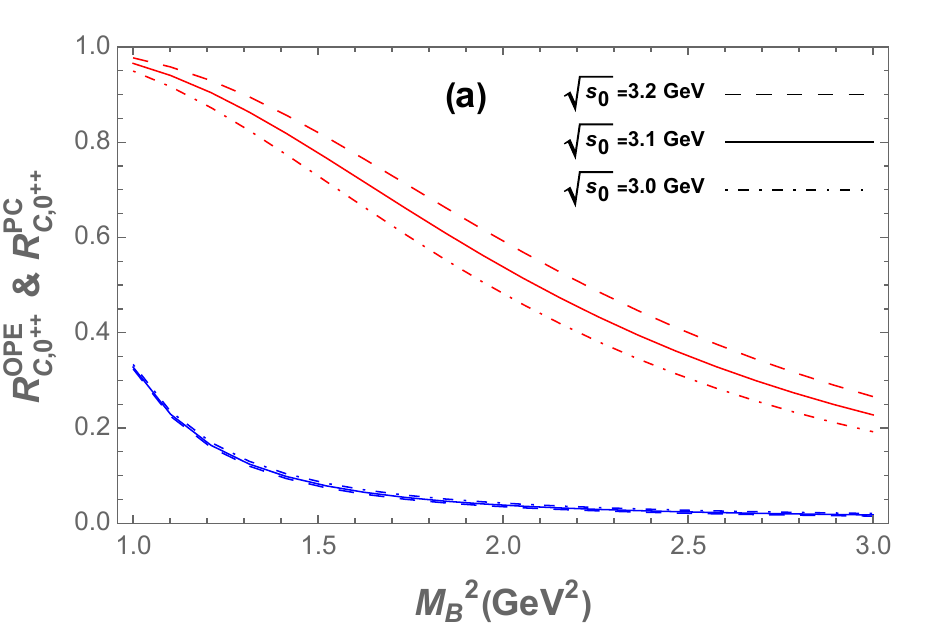}
\includegraphics[width=6.8cm]{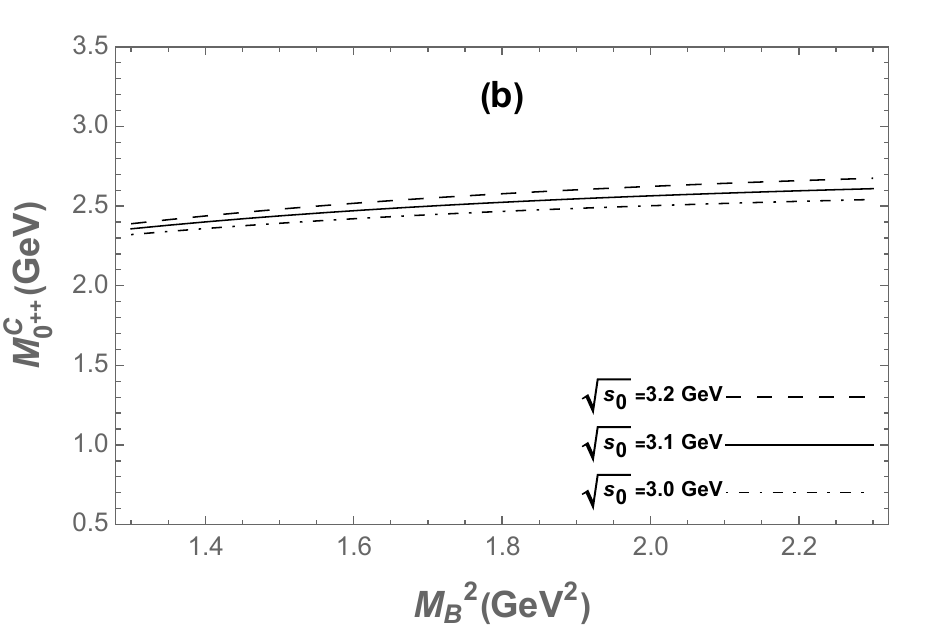}
\caption{ (a) The ratios ${R^{OPE}_{C,0^{++}}}$ and ${R^{PC}_{C,0^{++}}}$ as functions of the Borel parameter $M_B^2$ for different values of $\sqrt{s_0}$ for current (\ref{Jc0++}), where blue lines represent ${R^{OPE}_{C,0^{++}}}$ and red lines denote ${R^{PC}_{C,0^{++}}}$. (b) The mass of $0^{++}$ fully strange tetraquark state as a function of the Borel parameter $M_B^2$ for different values of $\sqrt{s_0}$.} \label{figC0++}
\end{figure}

\begin{figure}
\includegraphics[width=6.8cm]{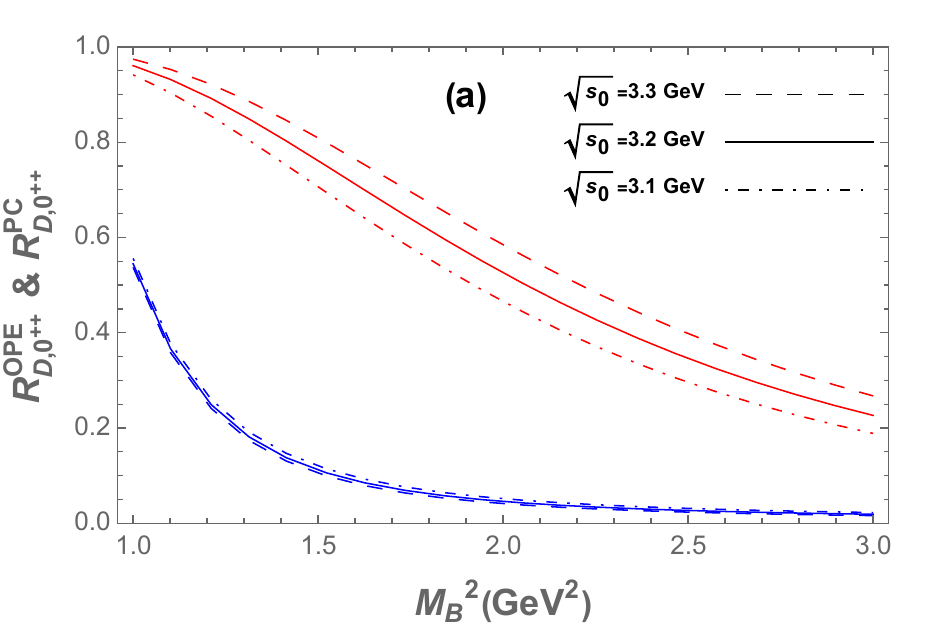}
\includegraphics[width=6.8cm]{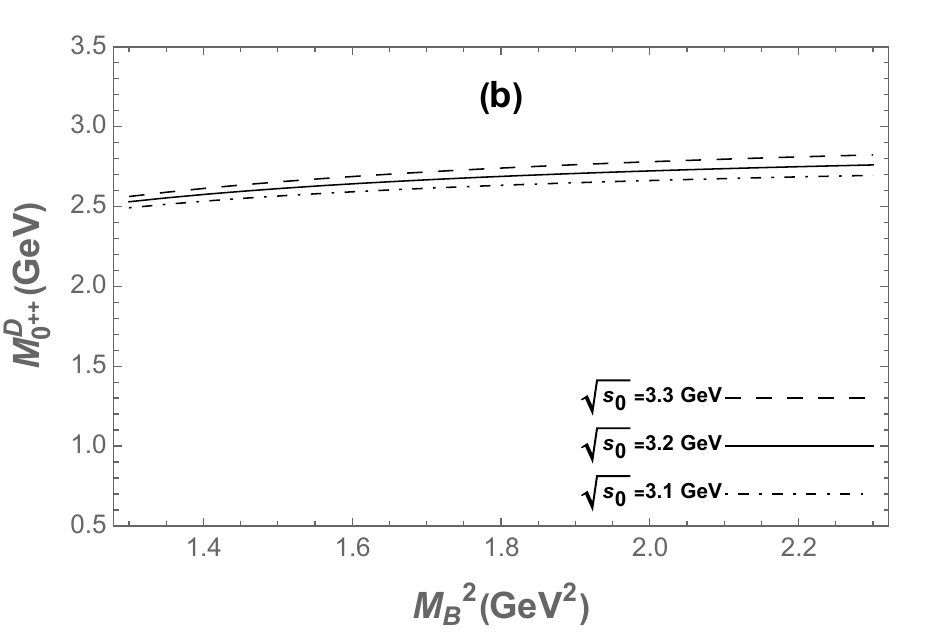}
\caption{ (a) The ratios ${R^{OPE}_{D,0^{++}}}$ and ${R^{PC}_{D,0^{++}}}$ as functions of the Borel parameter $M_B^2$ for different values of $\sqrt{s_0}$ for current (\ref{Jd0++}), where blue lines represent ${R^{OPE}_{D,0^{++}}}$ and red lines denote ${R^{PC}_{D,0^{++}}}$. (b) The mass of $0^{++}$ fully strange tetraquark state as a function of the Borel parameter $M_B^2$ for different values of $\sqrt{s_0}$.} \label{figD0++}
\end{figure}

\begin{figure}
\includegraphics[width=6.8cm]{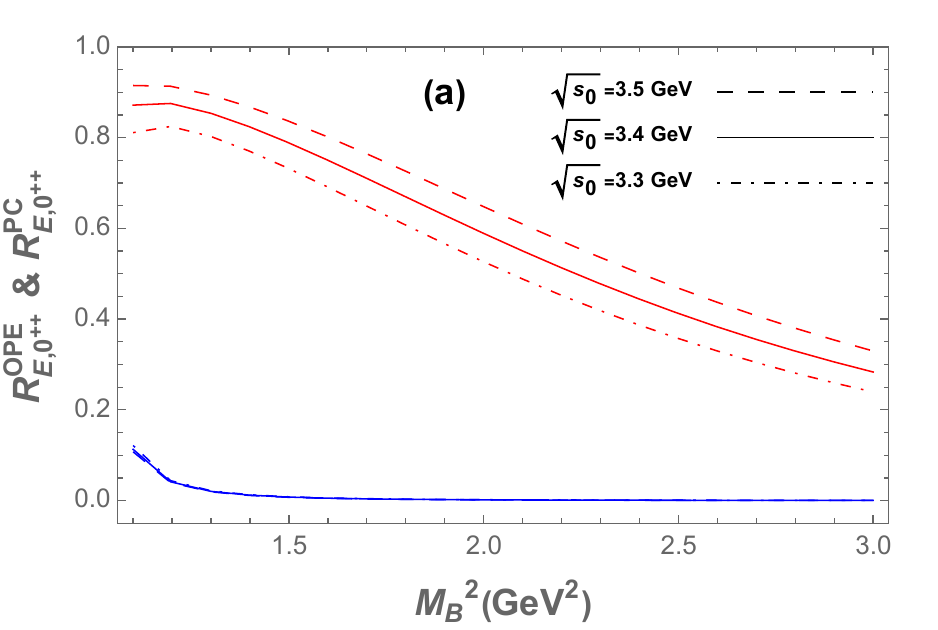}
\includegraphics[width=6.8cm]{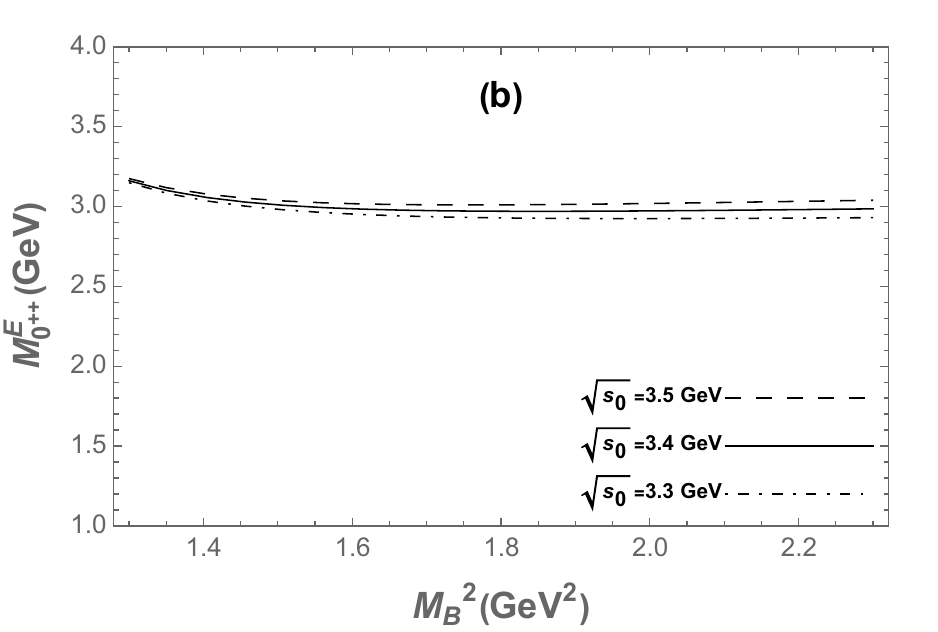}
\caption{ (a) The ratios ${R^{OPE}_{E,0^{++}}}$ and ${R^{PC}_{E,0^{++}}}$ as functions of the Borel parameter $M_B^2$ for different values of $\sqrt{s_0}$ for current (\ref{Je0++}), where blue lines represent ${R^{OPE}_{E,0^{++}}}$ and red lines denote ${R^{PC}_{E,0^{++}}}$. (b) The mass of $0^{++}$ fully strange tetraquark state as a function of the Borel parameter $M_B^2$ for different values of $\sqrt{s_0}$.} \label{figE0++}
\end{figure}

\begin{figure}
\includegraphics[width=6.8cm]{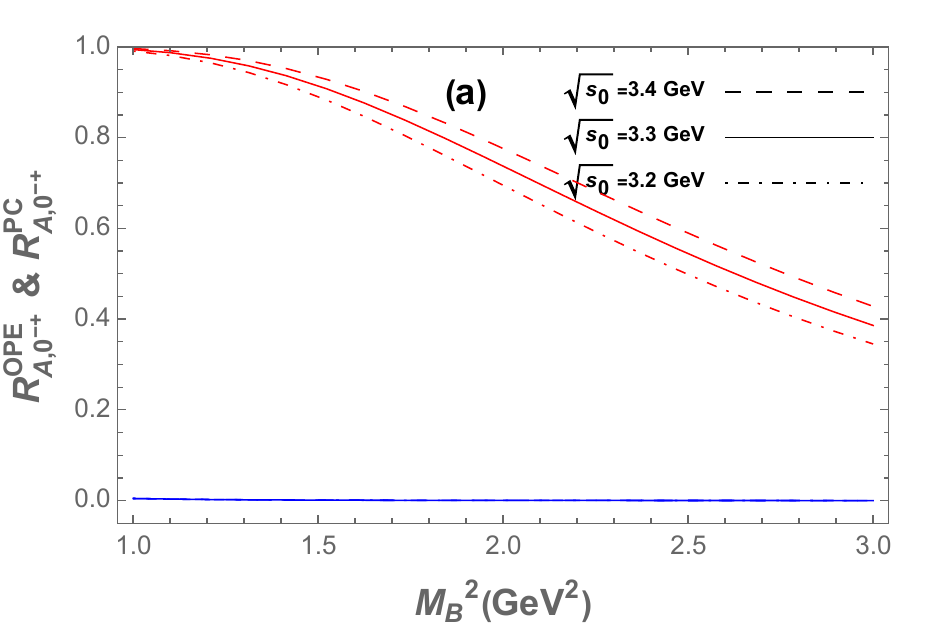}
\includegraphics[width=6.8cm]{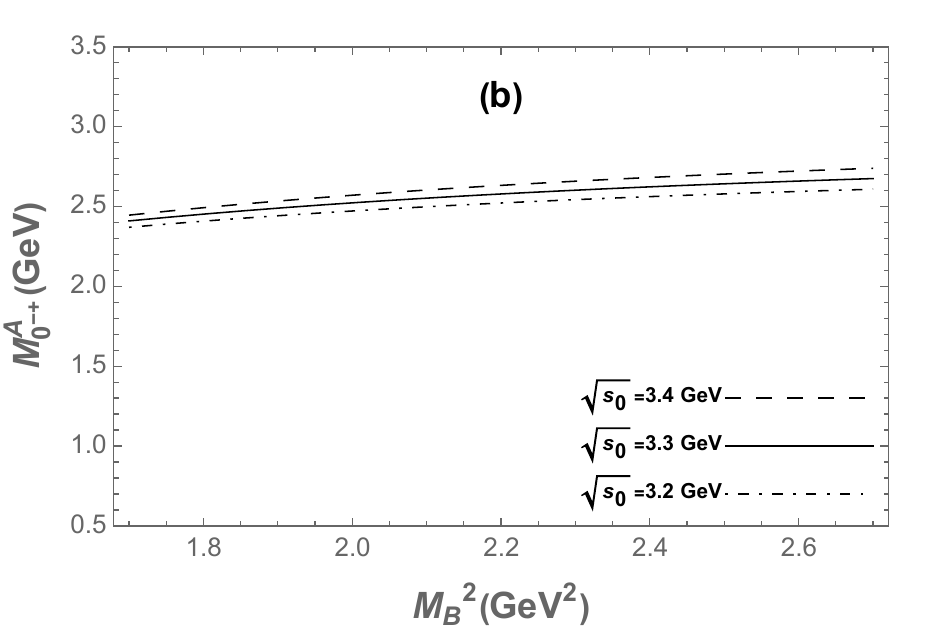}
\caption{ (a) The ratios ${R^{OPE}_{A,0^{-+}}}$ and ${R^{PC}_{A,0^{-+}}}$ as functions of the Borel parameter $M_B^2$ for different values of $\sqrt{s_0}$ for current (\ref{Ja0-+}), where blue lines represent ${R^{OPE}_{A,0^{-+}}}$ and red lines denote ${R^{PC}_{A,0^{-+}}}$. (b) The mass of $0^{-+}$ fully strange tetraquark state as a function of the Borel parameter $M_B^2$ for different values of $\sqrt{s_0}$.} \label{figA0-+}
\end{figure}

\begin{figure}
\includegraphics[width=6.8cm]{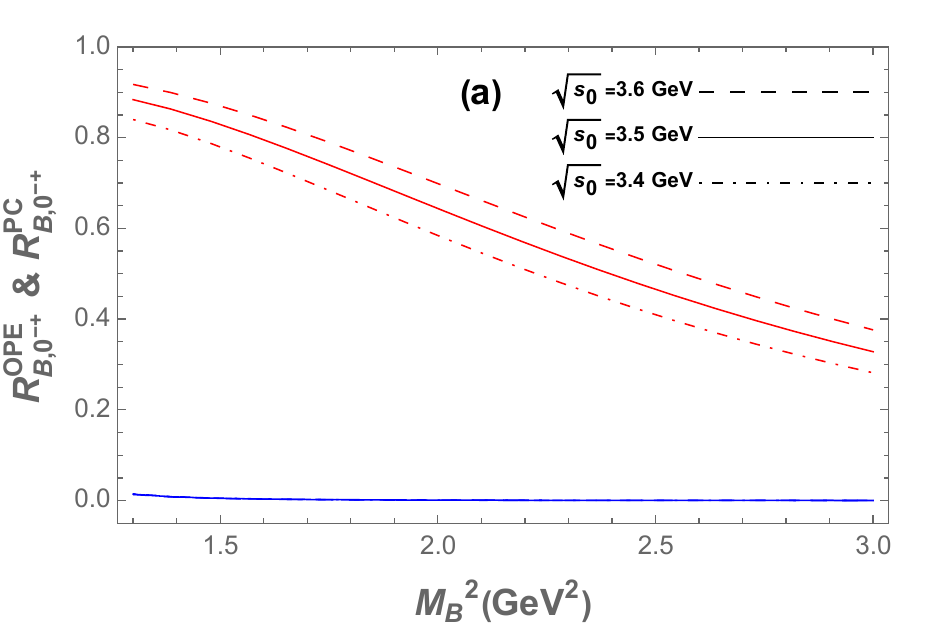}
\includegraphics[width=6.8cm]{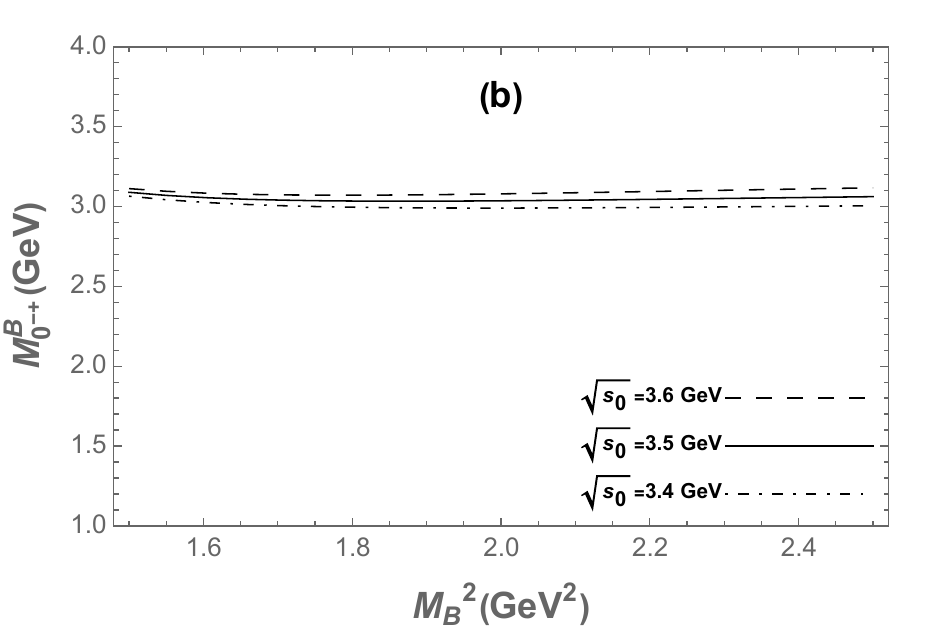}
\caption{ (a) The ratios ${R^{OPE}_{B,0^{-+}}}$ and ${R^{PC}_{B,0^{-+}}}$ as functions of the Borel parameter $M_B^2$ for different values of $\sqrt{s_0}$ for current (\ref{Jb0-+}), where blue lines represent ${R^{OPE}_{B,0^{-+}}}$ and red lines denote ${R^{PC}_{B,0^{-+}}}$. (b) The mass of $0^{-+}$ fully strange tetraquark state as a function of the Borel parameter $M_B^2$ for different values of $\sqrt{s_0}$.} \label{figB0-+}
\end{figure}

\begin{figure}
\includegraphics[width=6.8cm]{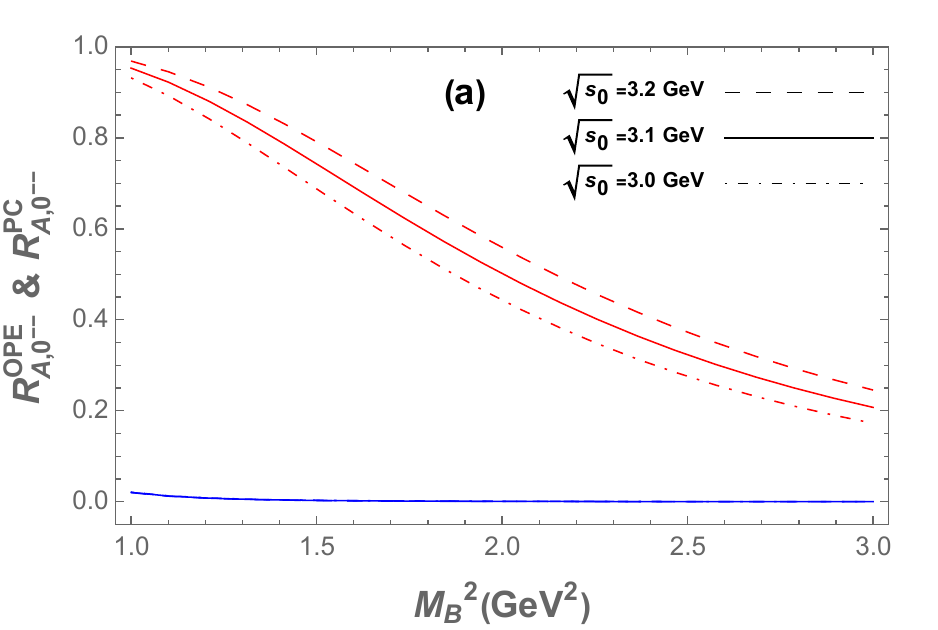}
\includegraphics[width=6.8cm]{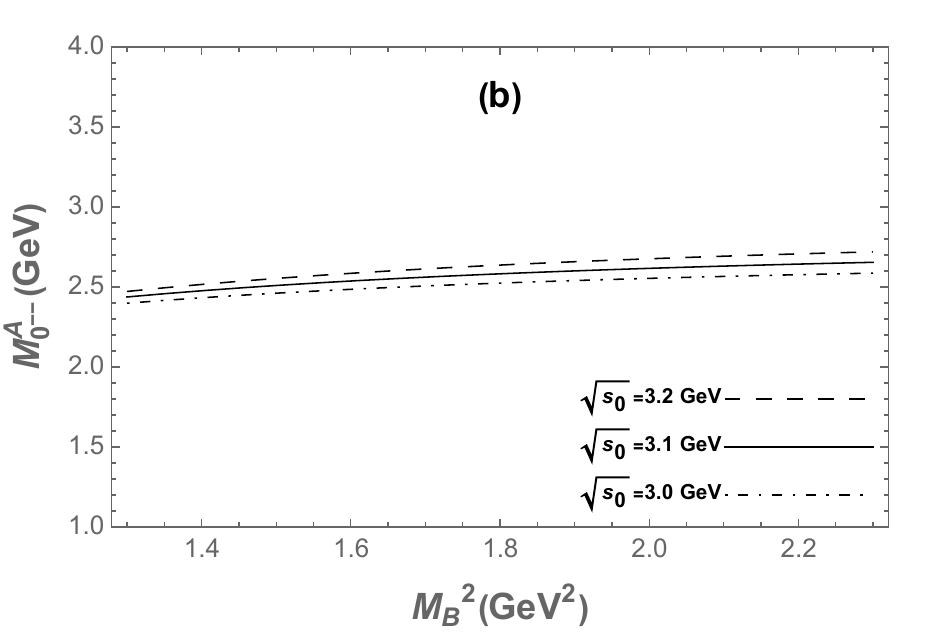}
\caption{ (a) The ratios ${R^{OPE}_{A,0^{--}}}$ and ${R^{PC}_{A,0^{--}}}$ as functions of the Borel parameter $M_B^2$ for different values of $\sqrt{s_0}$ for current (\ref{Ja0--}), where blue lines represent ${R^{OPE}_{A,0^{--}}}$ and red lines denote ${R^{PC}_{A,0^{--}}}$. (b) The mass of $0^{--}$ fully strange tetraquark state as a function of the Borel parameter $M_B^2$ for different values of $\sqrt{s_0}$.} \label{figA0--}
\end{figure}

\begin{figure}
\includegraphics[width=6.8cm]{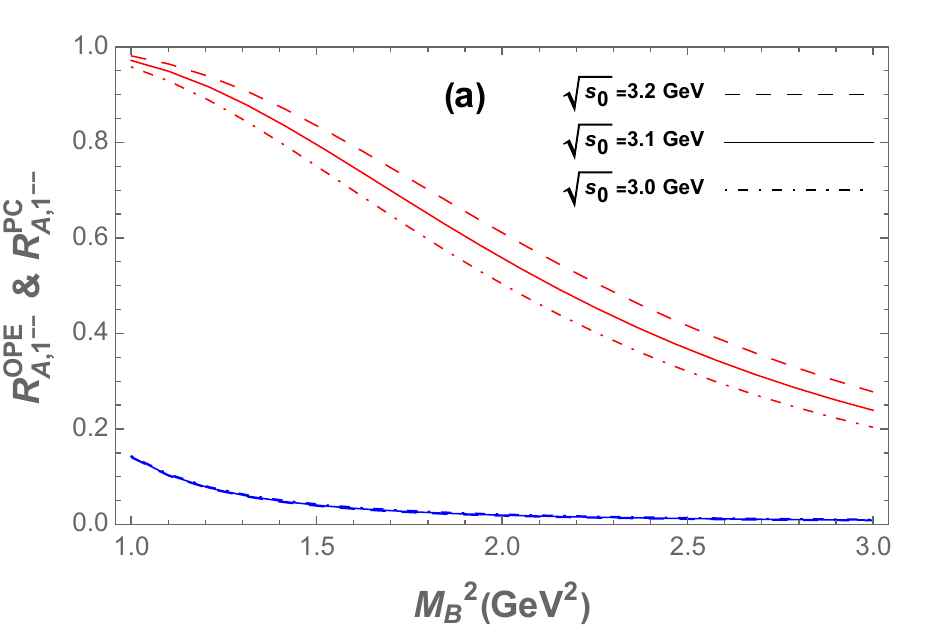}
\includegraphics[width=6.8cm]{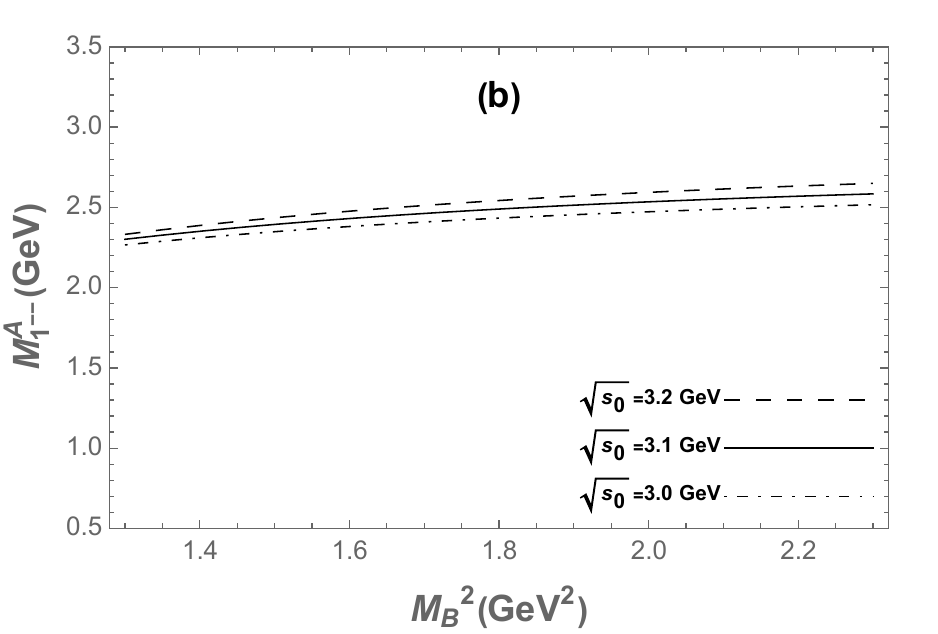}
\caption{ (a) The ratios ${R^{OPE}_{A,1^{--}}}$ and ${R^{PC}_{A,1^{--}}}$ as functions of the Borel parameter $M_B^2$ for different values of $\sqrt{s_0}$ for current (\ref{Ja1--}), where blue lines represent ${R^{OPE}_{A,1^{--}}}$ and red lines denote ${R^{PC}_{A,1^{--}}}$. (b) The mass of $1^{--}$ fully strange tetraquark state as a function of the Borel parameter $M_B^2$ for different values of $\sqrt{s_0}$.} \label{figA1--}
\end{figure}

\begin{figure}
\includegraphics[width=6.8cm]{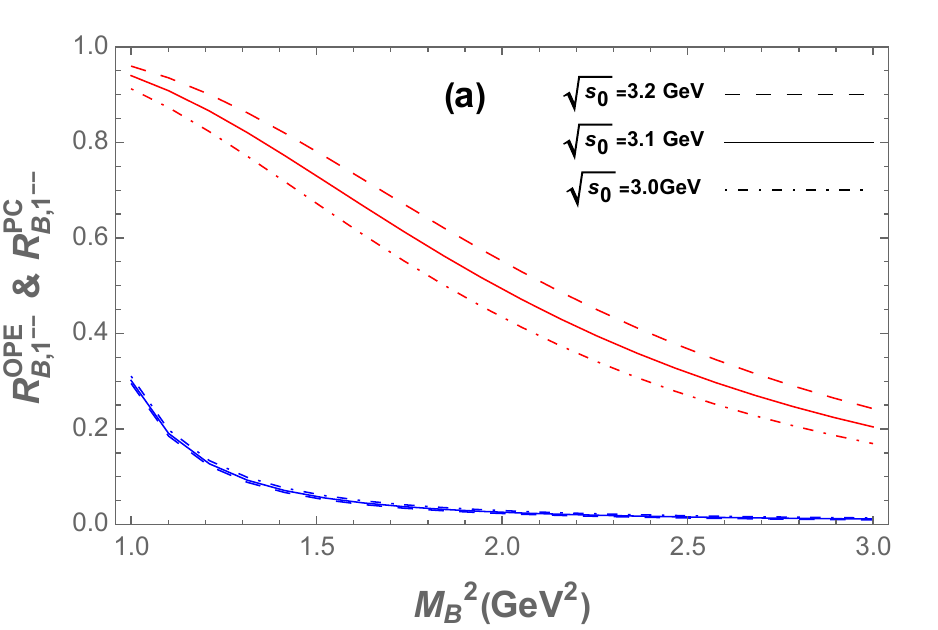}
\includegraphics[width=6.8cm]{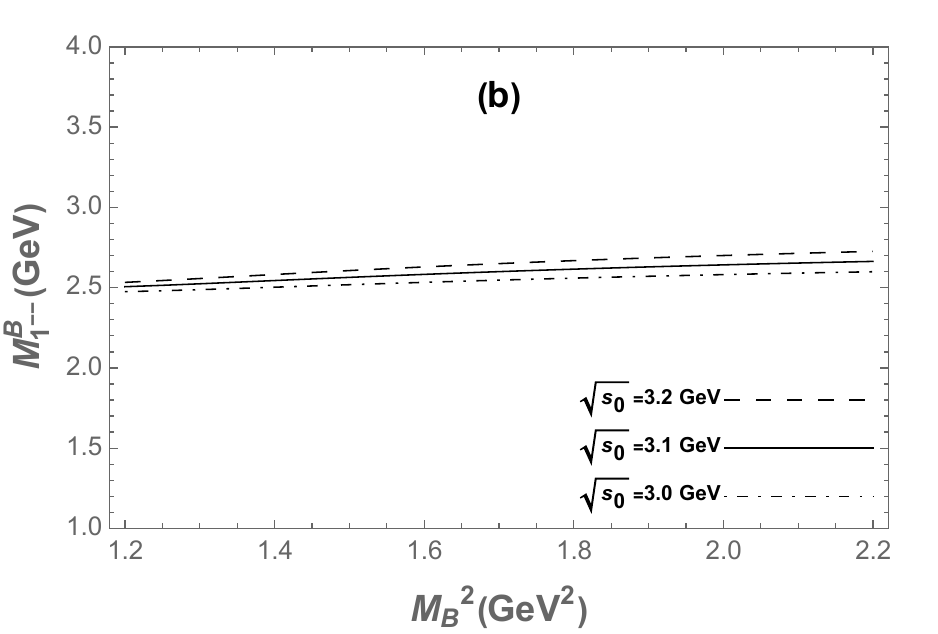}
\caption{ (a) The ratios ${R^{OPE}_{B,1^{--}}}$ and ${R^{PC}_{B,1^{--}}}$ as functions of the Borel parameter $M_B^2$ for different values of $\sqrt{s_0}$ for current (\ref{Jb1--}), where blue lines represent ${R^{OPE}_{B,1^{--}}}$ and red lines denote ${R^{PC}_{B,1^{--}}}$. (b) The mass of $1^{--}$ fully strange tetraquark state as a function of the Borel parameter $M_B^2$ for different values of $\sqrt{s_0}$.} \label{figB1--}
\end{figure}

\begin{figure}
\includegraphics[width=6.8cm]{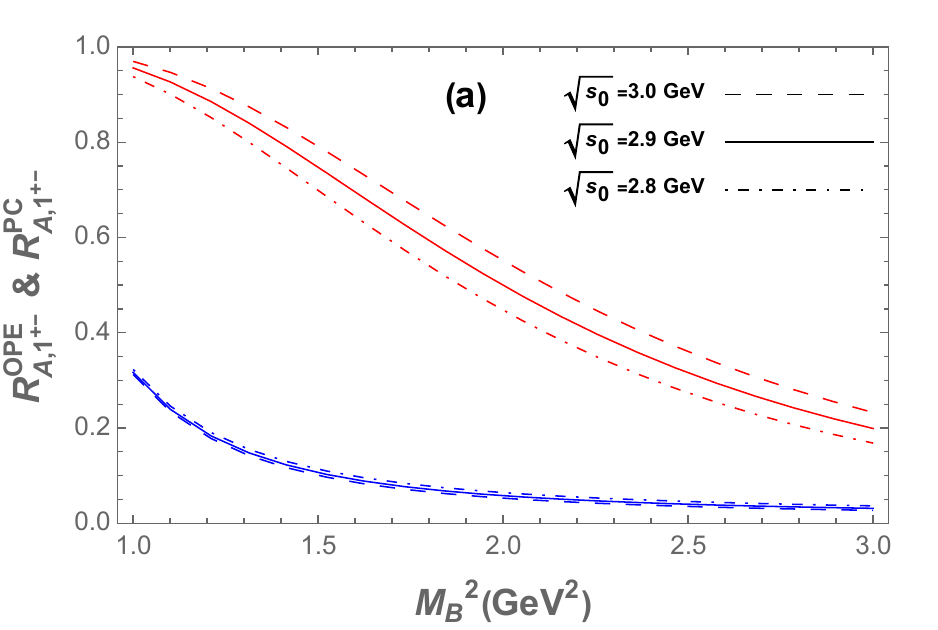}
\includegraphics[width=6.8cm]{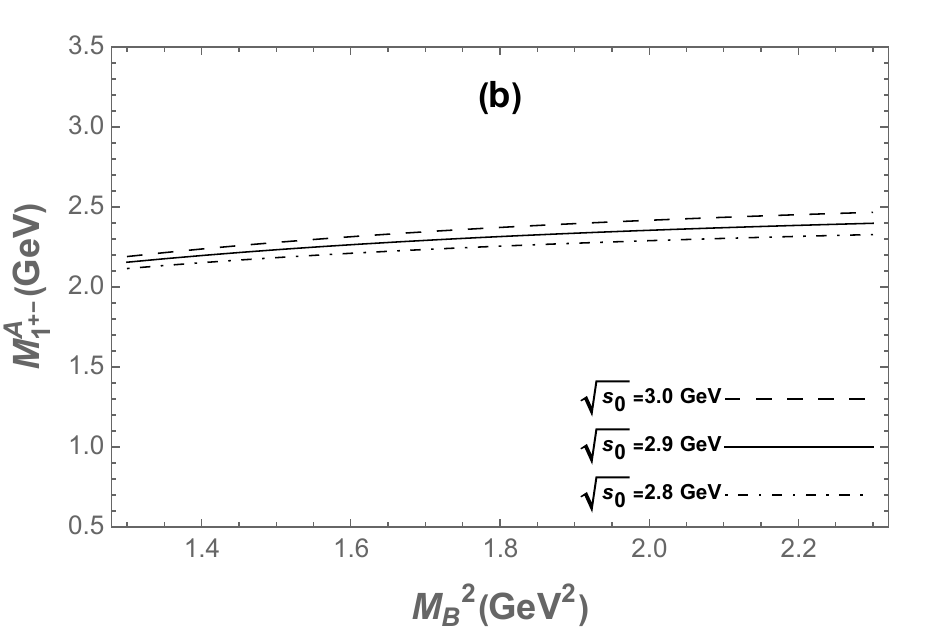}
\caption{ (a) The ratios ${R^{OPE}_{A,1^{+-}}}$ and ${R^{PC}_{A,1^{+-}}}$ as functions of the Borel parameter $M_B^2$ for different values of $\sqrt{s_0}$ for current (\ref{Ja1+-}), where blue lines represent ${R^{OPE}_{A,1^{+-}}}$ and red lines denote ${R^{PC}_{A,1^{+-}}}$. (b) The mass of $1^{+-}$ fully strange tetraquark state as a function of the Borel parameter $M_B^2$ for different values of $\sqrt{s_0}$.} \label{figA1+-}
\end{figure}

\begin{figure}
\includegraphics[width=6.8cm]{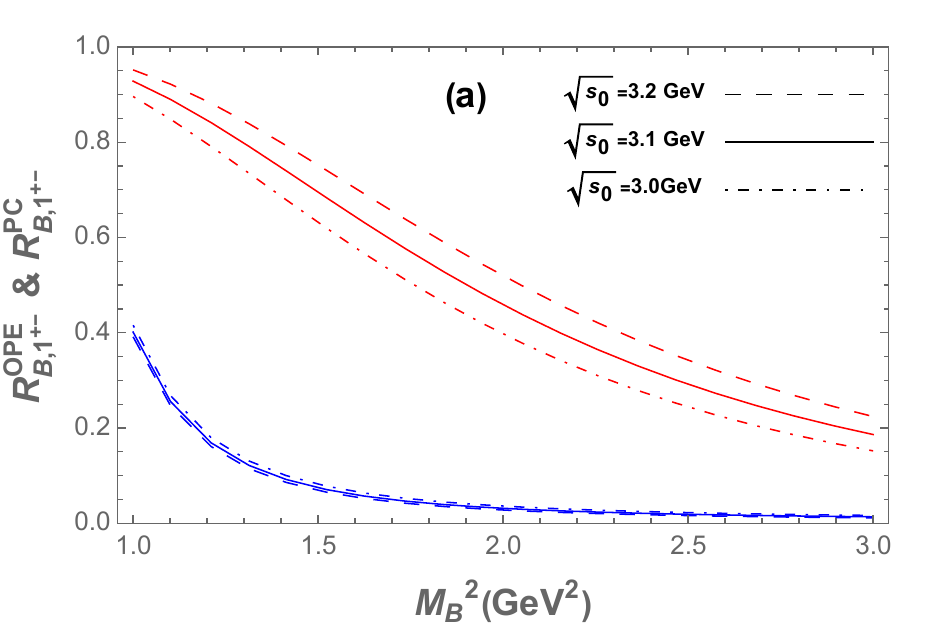}
\includegraphics[width=6.8cm]{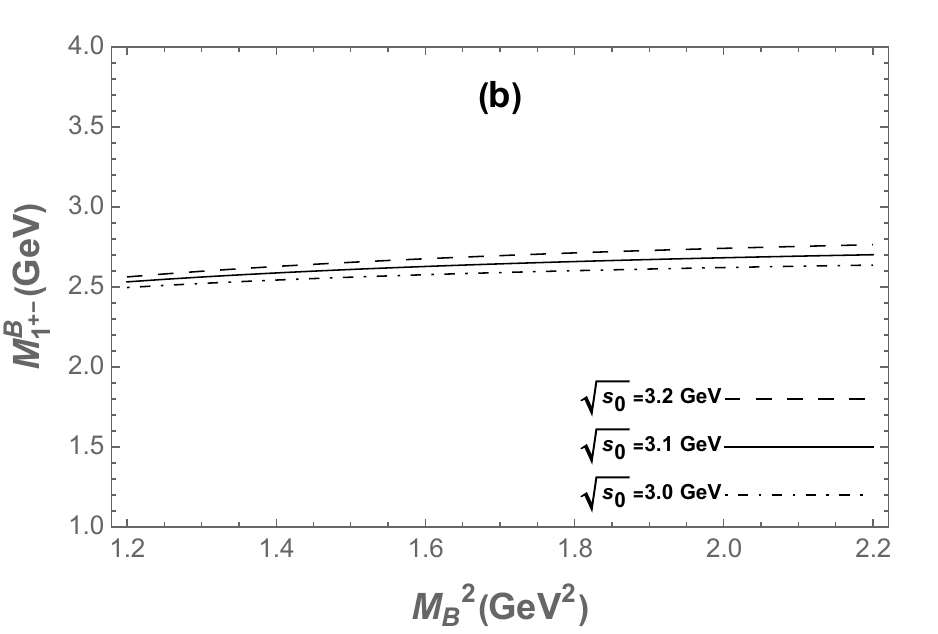}
\caption{ (a) The ratios ${R^{OPE}_{B,1^{+-}}}$ and ${R^{PC}_{B,1^{+-}}}$ as functions of the Borel parameter $M_B^2$ for different values of $\sqrt{s_0}$ for current (\ref{Jb1+-}), where blue lines represent ${R^{OPE}_{B,1^{+-}}}$ and red lines denote ${R^{PC}_{B,1^{+-}}}$. (b) The mass of $1^{+-}$ fully strange tetraquark state as a function of the Borel parameter $M_B^2$ for different values of $\sqrt{s_0}$.} \label{figB1+-}
\end{figure}

\begin{figure}
\includegraphics[width=6.8cm]{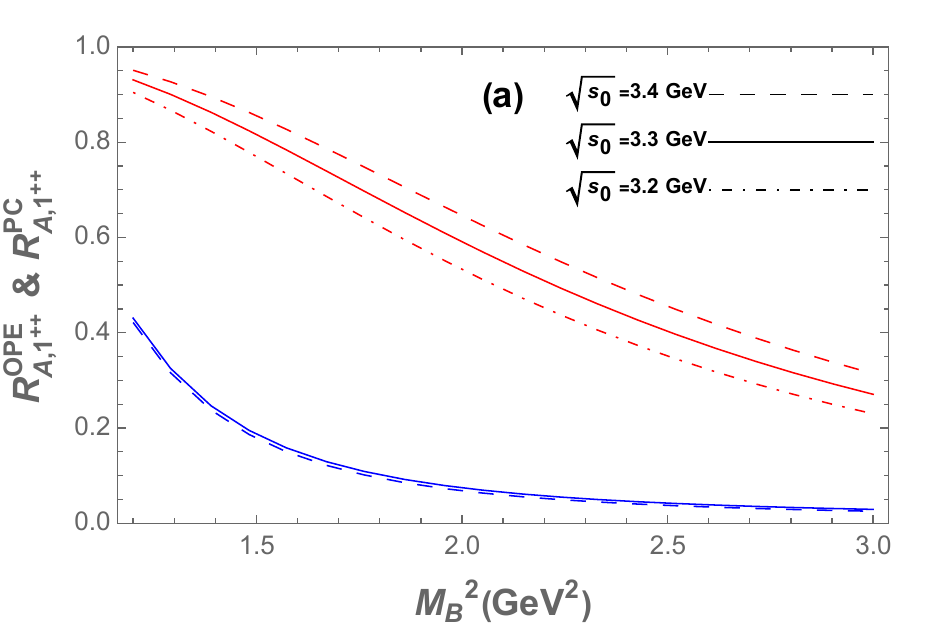}
\includegraphics[width=6.8cm]{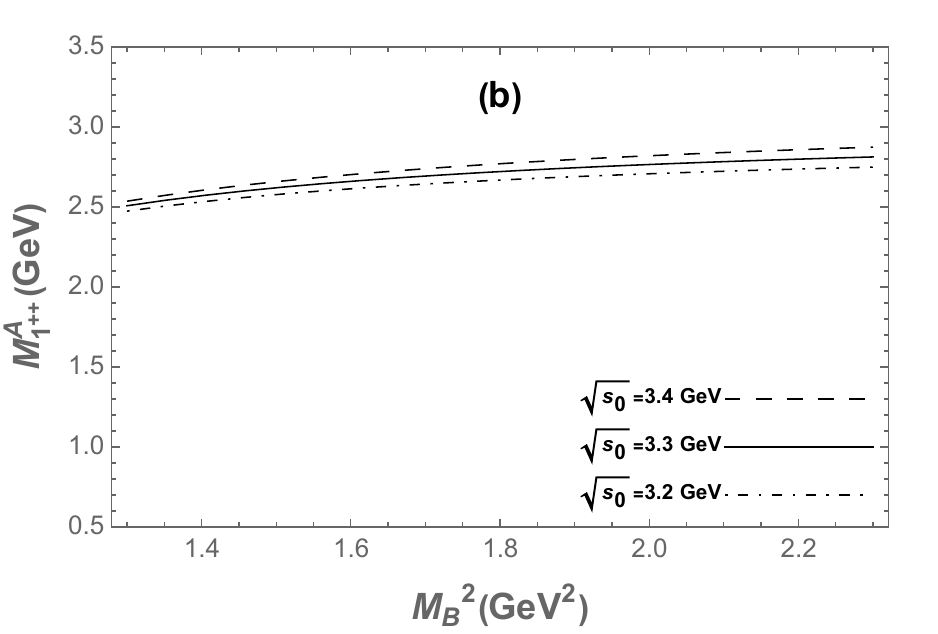}
\caption{ (a) The ratios ${R^{OPE}_{A,1^{++}}}$ and ${R^{PC}_{A,1^{++}}}$ as functions of the Borel parameter $M_B^2$ for different values of $\sqrt{s_0}$ for current (\ref{Ja1++}), where blue lines represent ${R^{OPE}_{A,1^{++}}}$ and red lines denote ${R^{PC}_{A,1^{++}}}$. (b) The mass of $1^{++}$ fully strange tetraquark state as a function of the Borel parameter $M_B^2$ for different values of $\sqrt{s_0}$.} \label{figA1++}
\end{figure}

\begin{figure}
\includegraphics[width=6.8cm]{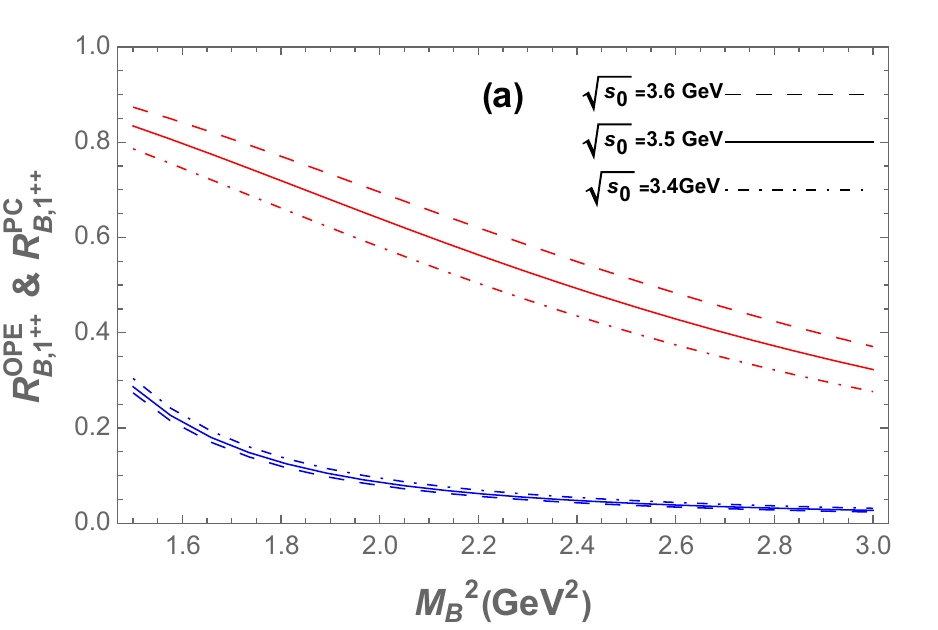}
\includegraphics[width=6.8cm]{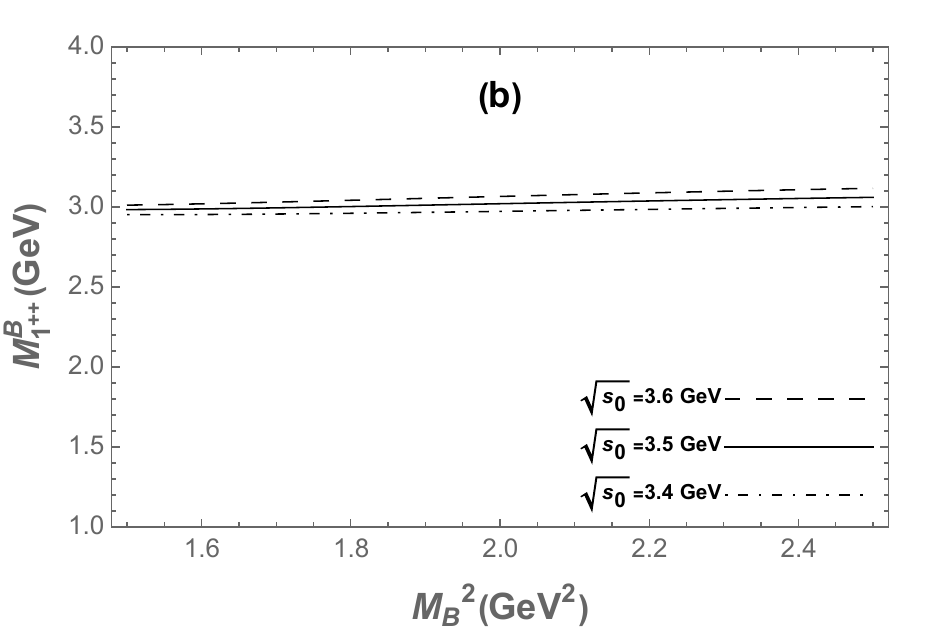}
\caption{ (a) The ratios ${R^{OPE}_{B,1^{++}}}$ and ${R^{PC}_{B,1^{++}}}$ as functions of the Borel parameter $M_B^2$ for different values of $\sqrt{s_0}$ for current (\ref{Jb1++}), where blue lines represent ${R^{OPE}_{B,1^{++}}}$ and red lines denote ${R^{PC}_{B,1^{++}}}$. (b) The mass of $1^{++}$ fully strange tetraquark state as a function of the Borel parameter $M_B^2$ for different values of $\sqrt{s_0}$.} \label{figB1++}
\end{figure}

\end{widetext}


\begin{thebibliography}{99}

%
\bibitem{GellMann:1964nj}
  M.~Gell-Mann,
  %``A Schematic Model of Baryons and Mesons,''
  Phys.\ Lett.\  {\bf 8}, 214 (1964).

\bibitem{Zweig}
  G.~Zweig, Report No. CERN-TH-401.

\bibitem{Choi:2003ue}
  S.~K.~Choi {\it et al.} [Belle Collaboration],
  %``Observation of a narrow charmonium - like state in exclusive B+- ---> K+- pi+ pi- J / psi decays,''
  Phys.\ Rev.\ Lett.\  {\bf 91}, 262001 (2003).
 %

 \bibitem{BESIII:2010gmv}
M.~Ablikim \textit{et al.} [BESIII],
%``Confirmation of the $X(1835)$ and observation of the resonances $X(2120)$ and $X(2370)$ in $J/\psi\to \gamma \pi^+\pi^-\eta^\prime$,''
Phys.\ Rev.\ Lett. {\bf 106}, 072002 (2011).
%doi:10.1103/PhysRevLett.106.072002
%[arXiv:1012.3510 [hep-ex]].
%148 citations counted in INSPIRE as of 15 Jul 2021

\bibitem{BES:2003aic}
J.~Z.~Bai \textit{et al.} [BES],
%``Observation of a near threshold enhancement in th p anti-p mass spectrum from radiative J / psi ---\ensuremath{>} gamma p anti-p decays,''
Phys.\ Rev.\ Lett. {\bf 91}, 022001 (2003).
%doi:10.1103/PhysRevLett.91.022001
%[arXiv:hep-ex/0303006 [hep-ex]].
%357 citations counted in INSPIRE as of 15 Jul 2021

\bibitem{BES:2005ega}
M.~Ablikim \textit{et al.} [BES],
%``Observation of a resonance X(1835) in J / psi -\ensuremath{>} gamma pi+ pi- eta-prime,''
Phys.\ Rev.\ Lett. {\bf 95}, 262001 (2005).
%doi:10.1103/PhysRevLett.95.262001
%[arXiv:hep-ex/0508025 [hep-ex]].
%206 citations counted in INSPIRE as of 15 Jul 2021

\bibitem{BESIII:2010vwa}
M.~Ablikim \textit{et al.} [BESIII],
%``Observation of a $p\bar{p}$ mass threshoud enhancement in $\psi^\prime\to\pi^+\pi^-J/\psi(J/\psi\to\gamma p\bar{p})$ decay,''
Chin.\ Phys.\ C {\bf 34}, 421 (2010).
%doi:10.1088/1674-1137/34/4/001
%[arXiv:1001.5328 [hep-ex]].
%61 citations counted in INSPIRE as of 15 Jul 2021

\bibitem{BESIII:2019wkp}
M.~Ablikim \textit{et al.} [BESIII],
%``Observation of $X(2370)$ and search for X(2120) in $J/\psi \rightarrow \gamma K{\bar{K}} \eta '$,''
Eur.\ Phys.\ J.\ C {\bf 80}, 746 (2020).
%doi:10.1140/epjc/s10052-020-8078-4
%[arXiv:1912.11253 [hep-ex]].
%6 citations counted in INSPIRE as of 15 Jul 2021

\bibitem{BESIII:2016qzq}
M.~Ablikim \textit{et al.} [BESIII],
%``Observation of pseudoscalar and tensor resonances in $J/\psi\to \gamma \phi \phi$,''
Phys.\ Rev.\ D {\bf 93}, 112011 (2016).
%doi:10.1103/PhysRevD.93.112011
%[arXiv:1602.01523 [hep-ex]].
%45 citations counted in INSPIRE as of 15 Jul 2021

\bibitem{BESIII:2020vtu}
M.~Ablikim \textit{et al.} [BESIII],
%``Observation of a Resonant Structure in $e^{+}e^{-} \to K^{+}K^{-}\pi^{0}\pi^{0}$,''
Phys.\ Rev.\ Lett. {\bf 124}, 112001 (2020).
%doi:10.1103/PhysRevLett.124.112001
%[arXiv:2001.04131 [hep-ex]].
%18 citations counted in INSPIRE as of 15 Jul 2021

\bibitem{BESIII:2017kqw}
M.~Ablikim \textit{et al.} [BESIII],
%``Study of $J/\psi$ and $\psi(3686)$ decay to $\Lambda\bar{\Lambda}$ and $\Sigma^0\bar{\Sigma}^0$ final states,''
Phys.\ Rev.\ D {\bf 95}, 052003 (2017).
%doi:10.1103/PhysRevD.95.052003
%[arXiv:1701.07191 [hep-ex]].
%37 citations counted in INSPIRE as of 15 Jul 2021

\bibitem{BESIII:2017hyw}
M.~Ablikim \textit{et al.} [BESIII],
%``Observation of a cross-section enhancement near mass threshold in $e^{+}e^{-}\rightarrow\Lambda\bar{\Lambda}$,''
Phys.\ Rev.\ D {\bf 97}, 032013 (2018).
%doi:10.1103/PhysRevD.97.032013
%[arXiv:1709.10236 [hep-ex]].
%40 citations counted in INSPIRE as of 15 Jul 2021

\bibitem{BESIII:2019cuv}
M.~Ablikim \textit{et al.} [BESIII],
%``Measurement of the cross section for $e^{+}e^{-}\rightarrow\Xi^{-}\bar\Xi^{+}$ and observation of an excited $\Xi$ baryon,''
Phys.\ Rev.\ Lett. {\bf 124}, 032002 (2020).
%doi:10.1103/PhysRevLett.124.032002
%[arXiv:1910.04921 [hep-ex]].
%6 citations counted in INSPIRE as of 15 Jul 2021

%\cite{BESIII:2025wpp}
\bibitem{BESIII:2025wpp}
M.~Ablikim \textit{et al.} [BESIII],
%``Observation of an Axial-Vector State in the Study of the Decay {\ensuremath{\psi}}(3686){\textrightarrow}{\ensuremath{\phi}}{\ensuremath{\eta}}{\ensuremath{\eta}}',''
Phys. Rev. Lett. \textbf{134}, 191901 (2025).
%doi:10.1103/PhysRevLett.134.191901
%0 citations counted in INSPIRE as of 03 Jul 2025

%\cite{Li:2020xzs}
\bibitem{Li:2020xzs}
Q.~Li, L.~C.~Gui, M.~S.~Liu, Q.~F.~L{\"u} and X.~H.~Zhong,
%``Mass spectrum and strong decays of strangeonium in a constituent quark model,''
Chin. Phys. C \textbf{45}, 023116 (2021).
%doi:10.1088/1674-1137/abcf22
%[arXiv:2004.05786 [hep-ph]].
%46 citations counted in INSPIRE as of 06 Jul 2025

%\cite{Xiao:2019qhl}
\bibitem{Xiao:2019qhl}
L.~Y.~Xiao, X.~Z.~Weng, X.~H.~Zhong and S.~L.~Zhu,
%``A possible explanation of the threshold enhancement in the process $e^+e^-\rightarrow \Lambda\bar{\Lambda}$,''
Chin. Phys. C \textbf{43}, 113105 (2019).
%doi:10.1088/1674-1137/43/11/113105
%[arXiv:1904.06616 [hep-ph]].
%24 citations counted in INSPIRE as of 06 Jul 2025

%\cite{Ishida:1986vn}
\bibitem{Ishida:1986vn}
S.~Ishida and K.~Yamada,
%``Light Quark Meson Spectrum in the Covariant Oscillator Quark Model With One Gluon Exchange Effects,''
Phys. Rev. D \textbf{35}, 265 (1987).
%doi:10.1103/PhysRevD.35.265
%46 citations counted in INSPIRE as of 06 Jul 2025

%\cite{Ebert:2009ub}
\bibitem{Ebert:2009ub}
D.~Ebert, R.~N.~Faustov and V.~O.~Galkin,
%``Mass spectra and Regge trajectories of light mesons in the relativistic quark model,''
Phys. Rev. D \textbf{79}, 114029 (2009).
%doi:10.1103/PhysRevD.79.114029
%[arXiv:0903.5183 [hep-ph]].
%280 citations counted in INSPIRE as of 06 Jul 2025

%\cite{Oudichhya:2023lva}
\bibitem{Oudichhya:2023lva}
J.~Oudichhya, K.~Gandhi and A.~K.~Rai,
%``Kaon and strangeonium spectrum in Regge phenomenology,''
Phys. Rev. D \textbf{108}, 014034 (2023).
%doi:10.1103/PhysRevD.108.014034
%[arXiv:2307.09012 [hep-ph]].
%19 citations counted in INSPIRE as of 06 Jul 2025

%\cite{Wang:2019qyy}
\bibitem{Wang:2019qyy}
L.~M.~Wang, J.~Z.~Wang, S.~Q.~Luo, J.~He and X.~Liu,
%``Studying $X(2100)$ hadronic decays and predicting its pion and kaon induced productions,''
Phys. Rev. D \textbf{101}, 034021 (2020).
%doi:10.1103/PhysRevD.101.034021
%[arXiv:1901.00636 [hep-ph]].
%19 citations counted in INSPIRE as of 06 Jul 2025

%\cite{Chen:2015iqa}
\bibitem{Chen:2015iqa}
K.~Chen, C.~Q.~Pang, X.~Liu and T.~Matsuki,
%``Light axial vector mesons,''
Phys. Rev. D \textbf{91}, 074025 (2015).
%doi:10.1103/PhysRevD.91.074025
%[arXiv:1501.07766 [hep-ph]].
%46 citations counted in INSPIRE as of 06 Jul 2025

%\cite{Liu:2020lpw}
\bibitem{Liu:2020lpw}
F.~X.~Liu, M.~S.~Liu, X.~H.~Zhong and Q.~Zhao,
%``Fully-strange tetraquark $ss\bar{s}\bar{s}$ spectrum and possible experimental evidence,''
Phys. Rev. D \textbf{103}, 016016 (2021).
%doi:10.1103/PhysRevD.103.016016
%[arXiv:2008.01372 [hep-ph]].
%33 citations counted in INSPIRE as of 06 Jul 2025

%\cite{Su:2022eun}
\bibitem{Su:2022eun}
N.~Su and H.~X.~Chen,
%``S- and P-wave fully strange tetraquark states from QCD sum rules,''
Phys. Rev. D \textbf{106}, 014023 (2022).
%doi:10.1103/PhysRevD.106.014023
%[arXiv:2204.13959 [hep-ph]].
%26 citations counted in INSPIRE as of 08 Dec 2025

%\cite{Xi:2023byo}
\bibitem{Xi:2023byo}
H.~Z.~Xi, Y.~W.~Jiang, H.~X.~Chen, A.~Hosaka and N.~Su,
%``Fully-strange tetraquark states with the exotic quantum numbers $J^{PC} = 0^{+-}$ and $2^{+-}$,''
Phys. Rev. D \textbf{108}, 094019 (2023).
%doi:10.1103/PhysRevD.108.094019
%[arXiv:2307.07819 [hep-ph]].
%8 citations counted in INSPIRE as of 08 Dec 2025

%\cite{Patel:2025bdu}
\bibitem{Patel:2025bdu}
V.~Patel, J.~Oudichhya and A.~K.~Rai,
%``Mass spectra of $qq\bar{q}\bar{q}$, $ss\bar{s}\bar{s}$ and $qq\bar{s}\bar{s}$ tetraquarks using Regge phenomenology,''
Eur. Phys. J. A \textbf{61}, 218 (2025).
%doi:10.1140/epja/s10050-025-01701-7
%[arXiv:2510.04496 [hep-ph]].
%0 citations counted in INSPIRE as of 08 Dec 2025

%\cite{Dong:2022otb}
\bibitem{Dong:2022otb}
R.~R.~Dong, N.~Su and H.~X.~Chen,
%``Highly excited and exotic fully-strange tetraquark states,''
Eur. Phys. J. C \textbf{82}, 983 (2022).
%doi:10.1140/epjc/s10052-022-10955-0
%[arXiv:2206.09517 [hep-ph]].
%10 citations counted in INSPIRE as of 08 Dec 2025

%\cite{Dong:2023evc}
\bibitem{Dong:2023evc}
R.~R.~Dong, N.~Su, H.~X.~Chen and E.~L.~Cui,
%``QCD sum rule study on the fully strange tetraquark states of J$^{PC}$ = 2$^{++}$,''
Front. in Phys. \textbf{11}, 1184103 (2023).
%doi:10.3389/fphy.2023.1184103
%3 citations counted in INSPIRE as of 08 Dec 2025

%\cite{Ma:2024vsi}
\bibitem{Ma:2024vsi}
Y.~Ma, W.~L.~Wu, L.~Meng, Y.~K.~Chen and S.~L.~Zhu,
%``Fully strange tetraquark resonant states as the cousins of X(6900),''
Phys. Rev. D \textbf{110}, 074026 (2024).
%doi:10.1103/PhysRevD.110.074026
%[arXiv:2408.00503 [hep-ph]].
%11 citations counted in INSPIRE as of 08 Dec 2025

%\cite{Xin:2022qnv}
\bibitem{Xin:2022qnv}
Q.~Xin and Z.~G.~Wang,
%``Fully-light vector tetraquark states with explicit P-wave via QCD sum rules*,''
Chin. Phys. C \textbf{48}, 033104 (2024).
%doi:10.1088/1674-1137/ad181c
%[arXiv:2211.14993 [hep-ph]].
%8 citations counted in INSPIRE as of 08 Dec 2025

\bibitem{Shifman}
  M.A. Shifman, A.I. Vainshtein and V.I. Zakharov,
  Nucl. Phys. {\bf B147}, 385 (1979); ibid, Nucl. Phys. {\bf B147},
  448 (1979).
  
  
%\cite{Wan:2022xkx}
\bibitem{Wan:2022xkx}
B.~D.~Wan, S.~Q.~Zhang and C.~F.~Qiao,
%``Possible structure of the newly found exotic state \ensuremath{\eta}1(1855),''
Phys. Rev. D \textbf{106}, 074003 (2022).
%doi:10.1103/PhysRevD.106.074003
%[arXiv:2203.14014 [hep-ph]].
%21 citations counted in INSPIRE as of 01 Jun 2024

\bibitem{Albuquerque:2013ija}
  R.~M.~Albuquerque,
  arXiv:1306.4671 [hep-ph].

%\cite{Wang:2013vex}
\bibitem{Wang:2013vex}
Z.~G.~Wang and T.~Huang,
%``Analysis of the $X(3872)$, $Z_c(3900)$ and $Z_c(3885)$ as axial-vector tetraquark states with QCD sum rules,''
Phys. Rev. D \textbf{89}, 054019 (2014).
%doi:10.1103/PhysRevD.89.054019
%[arXiv:1310.2422 [hep-ph]].
%172 citations counted in INSPIRE as of 13 May 2024

%\cite{Matheus:2006xi}
\bibitem{Matheus:2006xi}
R.~D.~Matheus, S.~Narison, M.~Nielsen and J.~M.~Richard,
%``Can the X(3872) be a 1++ four-quark state?,''
Phys. Rev. D \textbf{75}, 014005 (2007).
%doi:10.1103/PhysRevD.75.014005
%[arXiv:hep-ph/0608297 [hep-ph]].
%269 citations counted in INSPIRE as of 07 Jul 2025

%\cite{Cui:2011fj}
\bibitem{Cui:2011fj}
C.~Y.~Cui, Y.~L.~Liu and M.~Q.~Huang,
%``Investigating different structures of the $Z_{b}$(10610)  and  $Z_{b}$(10650),''
Phys. Rev. D \textbf{85}, 074014 (2012).
%doi:10.1103/PhysRevD.85.074014
%[arXiv:1107.1343 [hep-ph]].
%57 citations counted in INSPIRE as of 07 Jul 2025

%\cite{Tang:2019nwv}
\bibitem{Tang:2019nwv}
L.~Tang, B.~D.~Wan, K.~Maltman and C.~F.~Qiao,
%``Doubly Heavy Tetraquarks in QCD Sum Rules,''
Phys. Rev. D \textbf{101}, 094032 (2020).
%doi:10.1103/PhysRevD.101.094032
%[arXiv:1911.10951 [hep-ph]].
%45 citations counted in INSPIRE as of 01 Jun 2024

\bibitem{P.Col}
  P. Colangelo and A. Khodjamirian, in {\it At the frontier of
  particle physics / Handbook of QCD}, edited by M. Shifman (World
  Scientific, Singapore, 2001), arXiv:hep-ph/0010175.

\bibitem{Govaerts:1984hc}
J.~Govaerts, L.~J.~Reinders, H.~R.~Rubinstein and J.~Weyers,
%``Hybrid Quarkonia From {QCD} Sum Rules,''
Nucl. Phys. B \textbf{258}, 215-229 (1985).

  \bibitem{Reinders:1984sr}
  L.~J.~Reinders, H.~Rubinstein and S.~Yazaki,
  Phys.\ Rept.\  {\bf 127}, 1 (1985).

  \bibitem{Narison:1989aq}
  S.~Narison,
  World Sci.\ Lect.\ Notes Phys.\  {\bf 26} 1 (1989).
  
%\cite{Tang:2021zti}
\bibitem{Tang:2021zti}
C.~M.~Tang, Y.~C.~Zhao and L.~Tang,
%``Mass predictions of vector (1--) double-gluon heavy quarkonium hybrids from QCD sum rules,''
Phys. Rev. D \textbf{105}, 114004 (2022).
%doi:10.1103/PhysRevD.105.114004
%[arXiv:2111.07328 [hep-ph]].
%11 citations counted in INSPIRE as of 01 Jun 2024

\bibitem{Qiao:2014vva}
C.~F.~Qiao and L.~Tang,
%``Finding the $0^{--}$ Glueball,''
Phys. Rev. Lett. {\bf 113}, 221601 (2014).
%doi:10.1103/PhysRevLett.113.221601
%[arXiv:1408.3995 [hep-ph]].
%17 citations counted in INSPIRE as of 25 Nov 2020

%\cite{Qiao:2013dda}
\bibitem{Qiao:2013dda}
C.~F.~Qiao and L.~Tang,
%``Interpretation of $Z_c(4025)$ as the hidden charm tetraquark states via QCD Sum Rules,''
Eur. Phys. J. C \textbf{74}, 2810 (2014).
%doi:10.1140/epjc/s10052-014-2810-x
%[arXiv:1308.3439 [hep-ph]].
%65 citations counted in INSPIRE as of 22 Jan 2025

%\cite{Tang:2015twt}
\bibitem{Tang:2015twt}
L.~Tang and C.~F.~Qiao,
%``Mass spectra of $0^{+-}$, $1^{-+}$, and $2^{+-}$ exotic glueballs,''
Nucl. Phys. B \textbf{904}, 282-296 (2016).
%doi:10.1016/j.nuclphysb.2016.01.017
%[arXiv:1509.00305 [hep-ph]].
%27 citations counted in INSPIRE as of 22 Jan 2025

%\cite{Qiao:2013xca}
\bibitem{Qiao:2013xca}
C.~F.~Qiao and L.~Tang,
%``Molecular states with hidden charm and strange in QCD Sum Rules,''
EPL \textbf{107}, 31001 (2014).
%doi:10.1209/0295-5075/107/31001
%[arXiv:1309.7596 [hep-ph]].
%9 citations counted in INSPIRE as of 22 Jan 2025

%\cite{Wan:2019ake}
\bibitem{Wan:2019ake}
B.~D.~Wan, L.~Tang and C.~F.~Qiao,
%``Hidden-bottom and -charm hexaquark states in QCD sum rules,''
Eur. Phys. J. C \textbf{80}, 121 (2020).
%doi:10.1140/epjc/s10052-020-7701-8
%[arXiv:1912.12046 [hep-ph]].
%16 citations counted in INSPIRE as of 01 Jun 2024

%\cite{Tang:2024zvf}
\bibitem{Tang:2024zvf}
C.~M.~Tang, C.~G.~Duan and L.~Tang,
%``Fully charmed tetraquark states in $8_{[c\bar{c}]}\otimes 8_{[c\bar{c}]}$ color structure via QCD sum rules,''
Eur. Phys. J. C \textbf{84}, 743 (2024).
%doi:10.1140/epjc/s10052-024-13102-z
%[arXiv:2405.05042 [hep-ph]].
%6 citations counted in INSPIRE as of 22 Jan 2025

%\cite{Qiao:2013raa}
\bibitem{Qiao:2013raa}
C.~F.~Qiao and L.~Tang,
%``Estimating the mass of the hidden charm $1^+(1^{+})$ tetraquark state via QCD sum rules,''
Eur. Phys. J. C \textbf{74}, 3122 (2014).
%doi:10.1140/epjc/s10052-014-3122-x
%[arXiv:1307.6654 [hep-ph]].
%57 citations counted in INSPIRE as of 22 Jan 2025

%\cite{Tang:2024kmh}
\bibitem{Tang:2024kmh}
C.~M.~Tang, C.~G.~Duan, L.~Tang and C.~F.~Qiao,
%``A novel configuration of gluonic tetraquark state,''
Eur. Phys. J. C \textbf{85}, 396 (2025).
%doi:10.1140/epjc/s10052-025-14106-z
%[arXiv:2411.11433 [hep-ph]].
%3 citations counted in INSPIRE as of 21 Jun 2025

%\cite{Wan:2020oxt}
\bibitem{Wan:2020oxt}
B.~D.~Wan and C.~F.~Qiao,
%``About the exotic structure of $Z_{cs}$,''
Nucl. Phys. B \textbf{968}, 115450 (2021).
%doi:10.1016/j.nuclphysb.2021.115450
%[arXiv:2011.08747 [hep-ph]].
%57 citations counted in INSPIRE as of 01 Jun 2024

%\cite{Wan:2020fsk}
\bibitem{Wan:2020fsk}
B.~D.~Wan and C.~F.~Qiao,
%``Gluonic tetracharm configuration of $X (6900)$,''
Phys. Lett. B \textbf{817}, 136339 (2021).
%doi:10.1016/j.physletb.2021.136339
%[arXiv:2012.00454 [hep-ph]].
%70 citations counted in INSPIRE as of 14 Jan 2025

%\cite{Li:2024ctd}
\bibitem{Li:2024ctd}
S.~N.~Li and L.~Tang,
%``Spectrum of $[8]_{[c\bar{s}]} \otimes [8]_{[q \bar{q^\prime}]}$ systems with quantum numbers $J^{P}=0^\pm$ and $1^\pm$,''
[arXiv:2404.11145 [hep-ph]].
%0 citations counted in INSPIRE as of 01 Jun 2024

%\cite{Wan:2021vny}
\bibitem{Wan:2021vny}
B.~D.~Wan, S.~Q.~Zhang and C.~F.~Qiao,
%``Light baryonium spectrum,''
Phys. Rev. D \textbf{105}, 014016 (2022).
%doi:10.1103/PhysRevD.105.014016
%[arXiv:2109.07130 [hep-ph]].
%12 citations counted in INSPIRE as of 01 Jun 2024

%\cite{Zhao:2023imq}
\bibitem{Zhao:2023imq}
Y.~C.~Zhao, C.~M.~Tang and L.~Tang,
%``Mass predictions of triply heavy hybrid baryons via QCD sum rules,''
Eur. Phys. J. C \textbf{83}, 654 (2023).
%doi:10.1140/epjc/s10052-023-11825-z
%[arXiv:2303.15173 [hep-ph]].
%0 citations counted in INSPIRE as of 01 Jun 2024

%\cite{Zhang:2022obn}
\bibitem{Zhang:2022obn}
S.~Q.~Zhang, B.~D.~Wan, L.~Tang and C.~F.~Qiao,
%``Gluonic nature of the newly observed state X(2600),''
Phys. Rev. D \textbf{106}, 074010 (2022).
%doi:10.1103/PhysRevD.106.074010
%[arXiv:2206.13133 [hep-ph]].
%2 citations counted in INSPIRE as of 01 Jun 2024

%\cite{Wan:2022uie}
\bibitem{Wan:2022uie}
B.~D.~Wan and C.~F.~Qiao,
%``The tetra-heavy baryonium spectra,''
[arXiv:2208.14042 [hep-ph]].
%2 citations counted in INSPIRE as of 01 Jun 2024

%\cite{Yin:2021cbb}
\bibitem{Yin:2021cbb}
F.~H.~Yin, W.~Y.~Tian, L.~Tang and Z.~H.~Guo,
%``Determination of the up/down-quark mass within QCD sum rules in the scalar channel,''
Eur. Phys. J. C \textbf{81}, 818 (2021).
%doi:10.1140/epjc/s10052-021-09599-3
%[arXiv:2104.08847 [hep-ph]].
%2 citations counted in INSPIRE as of 01 Jun 2024

%\cite{Wan:2023epq}
\bibitem{Wan:2023epq}
B.~D.~Wan,
%``Mass spectra of $0^{--}$ and $0^{+-}$ hidden-heavy baryoniums,''
Eur. Phys. J. C \textbf{84}, 760 (2024).
%doi:10.1140/epjc/s10052-024-13126-5
%[arXiv:2311.13161 [hep-ph]].
%2 citations counted in INSPIRE as of 05 Sep 2024

%\cite{Tang:2016pcf}
\bibitem{Tang:2016pcf}
L.~Tang and C.~F.~Qiao,
%``Tetraquark States with Open Flavors,''
Eur. Phys. J. C \textbf{76}, 558 (2016).
%doi:10.1140/epjc/s10052-016-4436-7
%[arXiv:1603.04761 [hep-ph]].
%47 citations counted in INSPIRE as of 22 Jan 2025

%\cite{Yang:2020wkh}
\bibitem{Yang:2020wkh}
B.~C.~Yang, L.~Tang and C.~F.~Qiao,
%``Scalar fully-heavy tetraquark states $QQ^\prime {\bar{Q}} \bar{Q^\prime }$ in QCD sum rules,''
Eur. Phys. J. C \textbf{81}, 324 (2021).
%doi:10.1140/epjc/s10052-021-09096-7
%[arXiv:2012.04463 [hep-ph]].
%55 citations counted in INSPIRE as of 01 Jun 2024

%\cite{Wan:2024dmi}
\bibitem{Wan:2024dmi}
B.~D.~Wan, 
%``Interpretation of X(3960) as the hidden charm-strange tetraquark states via QCD sum rules,''
Nucl. Phys. B \textbf{1004}, 116538 (2024).
%doi:10.1016/j.nuclphysb.2024.116538
%0 citations counted in INSPIRE as of 01 Jun 2024

%\cite{Wan:2024fam}
\bibitem{Wan:2024fam}
B.~D.~Wan and H.~T.~Xu,
%``0$^{– –}$ hidden-heavy tetraquark states via QCD sum rules*,''
Chin. Phys. C \textbf{48}, 093103 (2024).
%doi:10.1088/1674-1137/ad53b7
%0 citations counted in INSPIRE as of 05 Sep 2024

%\cite{Wan:2024ykm}
\bibitem{Wan:2024ykm}
B.~D.~Wan and Y.~R.~Wang,
%``Possible structure of $T_{c\bar{s}0}(2900)^{++}$,''
Eur. Phys. J. A \textbf{60}, 179 (2024).
%doi:10.1140/epja/s10050-024-01388-2
%1 citations counted in INSPIRE as of 14 Jan 2025

%\cite{Wan:2024pet}
\bibitem{Wan:2024pet}
B.~D.~Wan and S.~Yang,
%``Gluonic hidden-charm tetraquark states,''
Eur. Phys. J. A \textbf{61}, 11 (2025).
%doi:10.1140/epja/s10050-024-01475-4
%[arXiv:2407.18672 [hep-ph]].
%2 citations counted in INSPIRE as of 21 Jun 2025

%\cite{Zhang:2024jvv}
\bibitem{Zhang:2024jvv}
W.~S.~Zhang and L.~Tang,
%``Investigating triply heavy tetraquark states through QCD sum rules,''
[arXiv:2412.11531 [hep-ph]].
%3 citations counted in INSPIRE as of 21 Jun 2025

%\cite{Zhang:2023nxl}
\bibitem{Zhang:2023nxl}
S.~Q.~Zhang and C.~F.~Qiao,
%``{\ensuremath{\Lambda}}c semileptonic decays,''
Phys. Rev. D \textbf{108}, 074017 (2023).
%doi:10.1103/PhysRevD.108.074017
%[arXiv:2307.05019 [hep-ph]].
%10 citations counted in INSPIRE as of 06 Jul 2025

%\cite{Zhang:2024ick}
\bibitem{Zhang:2024ick}
S.~Q.~Zhang, X.~H.~Zhang and C.~F.~Qiao,
%``Hyperon semileptonic decays in QCD sum rules,''
JHEP \textbf{06}, 122 (2024).
%doi:10.1007/JHEP06(2024)122
%[arXiv:2402.15088 [hep-ph]].
%5 citations counted in INSPIRE as of 06 Jul 2025

%\cite{Zhang:2024asb}
\bibitem{Zhang:2024asb}
S.~Q.~Zhang and C.~F.~Qiao,
%``Rare {\ensuremath{\Lambda}}c decays and new physics effects,''
Phys. Rev. D \textbf{110}, 114040 (2024).
%doi:10.1103/PhysRevD.110.114040
%[arXiv:2411.15857 [hep-ph]].
%1 citations counted in INSPIRE as of 06 Jul 2025

%\cite{Zhang:2024ulk}
\bibitem{Zhang:2024ulk}
X.~H.~Zhang, S.~Q.~Zhang and C.~F.~Qiao,
%``The spectra of $p\bar\Lambda$ and $p\bar\Sigma$~hexaquark states,''
Eur. Phys. J. C \textbf{85}, 693 (2025).
%doi:10.1140/epjc/s10052-025-14356-x
%[arXiv:2412.20150 [hep-ph]].
%1 citations counted in INSPIRE as of 06 Jul 2025

%\cite{Fu:2018ngx}
\bibitem{Fu:2018ngx}
Y.~C.~Fu, Z.~R.~Huang, Z.~F.~Zhang and W.~Chen,
%``Exotic tetraquark states with $J^{PC}=0^{+-}$,''
Phys. Rev. D \textbf{99}, 014025 (2019).
%doi:10.1103/PhysRevD.99.014025
%[arXiv:1811.03333 [hep-ph]].
%10 citations counted in INSPIRE as of 17 Jul 2025

%\cite{Huang:2016rro}
\bibitem{Huang:2016rro}
Z.~R.~Huang, W.~Chen, T.~G.~Steele, Z.~F.~Zhang and H.~Y.~Jin,
%``Investigation of the light four-quark states with exotic $J^{PC}=0^{--}$,''
Phys. Rev. D \textbf{95}, 076017 (2017).
%doi:10.1103/PhysRevD.95.076017
%[arXiv:1610.02081 [hep-ph]].
%25 citations counted in INSPIRE as of 17 Jul 2025

%\cite{Jafarzade:2025qvx}
\bibitem{Jafarzade:2025qvx}
S.~Jafarzade and R.~F.~Lebed,
%``Hidden-Strangeness Tetraquarks in the Dynamical Diquark Model,''
[arXiv:2505.15704 [hep-ph]].
%1 citations counted in INSPIRE as of 17 Jul 2025

\end{thebibliography}
\end{document}